\documentclass[a4paper,11pt]{article}

\usepackage{geometry}
\usepackage{bm}
\usepackage{amsmath,amssymb}
\usepackage{graphicx}
\usepackage[noend]{algpseudocode}
\usepackage{algorithmicx,algorithm}
\usepackage[comma,authoryear]{natbib} 
\usepackage[labelfont=bf]{caption}
\usepackage{float}
\usepackage{multicol,lipsum}
\usepackage[switch,columnwise]{lineno}

\usepackage[T1]{fontenc}
\usepackage[utf8]{inputenc}
\usepackage{authblk}
\usepackage[colorlinks,linkcolor=blue,citecolor=blue,urlcolor=blue]{hyperref}
\usepackage[title]{appendix}
\usepackage[]{pifont}
\usepackage{lettrine}
\usepackage{hyperref}
\usepackage{setspace}

\graphicspath{{figures/}} 

\numberwithin{equation}{section}

\title{\textbf{{\LARGE Robust probabilistic measurement of structural-functional module consistency in infant brain development}}}

\author[a]{\small Lingbin Bian }
\author[a]{\small Feihong Liu} 
\author[a,c]{\small Qian Wang }
\author[a,c]{\small Han Zhang }
\author[a,b,c,*]{\small Dinggang Shen }
\author[ ]{\small the UNC/UMN Baby Connectome Project Consortium }

\affil[a]{School of Biomedical Engineering \& State Key Laboratory of Advanced Medical Materials and Devices, ShanghaiTech University, Shanghai 201210, China}
\affil[b]{Shanghai United Imaging Intelligence Co., Ltd., Shanghai 200230, China}
\affil[c]{Shanghai Clinical Research and Trial Center, Shanghai 201210, China}

\affil[*]{\small Corresponding authors: Han Zhang  (zhanghan2@shanghaitech.edu.cn), Dinggang Shen (Dinggang.Shen@gmail.com)}
\date{} 

\begin{document}
\renewcommand{\figurename}{\textbf{Fig.}} %
\maketitle


\renewcommand{\baselinestretch}{1.5}
\begin{abstract}
\setlength{\parindent}{0pt} \setlength{\parskip}{1.5ex plus 0.5ex
minus 0.2ex} 

Brain network is commonly divided into modules for analyzing their functionally segregated roles for group-level analysis in neuroimaging studies. Here, we introduce stochastic modules within brain networks for a robust probabilistic measurement of structural-functional module consistency (SFMC) in a group of subjects. Specifically, a stochastic module can be regarded as the chance of a brain region across subjects potentially being assigned to a group-level sub-network, characterized as an assignment probability for this brain region. This novel method has two advantages for evaluating inhomogeneous modules in brain networks. The first is that it can robustly evaluate the consistency between brain structural and functional modules whose population sizes are not necessary the same, and the second is that it is able to take into account the inter-individual variability of the modules for the groups. Moreover, compared with the conventional structural-functional coupling approach, our stochastic module-based method reveals a more pronounced decline in the coupling between structure and function, indicating stronger developmental reorganization. Our results using the dataset from Baby Connectome Project (BCP) show that the SFMC decreases from 0 to 5 years old, and is greater in primary brain regions, such as visual areas, while lower in more advanced cognitive regions, including those related to attention, control, and  default mode network.
\end{abstract}

\section{Introduction}
During infancy, the structural and functional networks of the brain undergo significant changes in specialization and interaction because of rapid brain development, which is usually characterized by modular structure as a fundamental feature of complex biological systems. A network module is a group of nodes (brain regions) that are tightly connected, usually forming specialized functions responsible for processing particular types of information \citep{Bian2025}. Across development from the neonatal period to adolescence, structural connectivity (SC) becomes increasingly modular and resilient to disruption, characterized by increased intra-module connectivity and decreased inter-module connectivity \citep{Huang2013,Baum2017}. In parallel with structural development, functional connectivity (FC) also shows enhanced modular organization \citep{Wen2019}, supporting more efficient brain network communication. However, it remains unclear how the SC and FC modules are consistent with each other during infant brain development, especially as higher-order cognitive functions become more developed and key behavioral milestones begin to emerge. 

It is essential to account for inter-individual variability in structural and functional networks when examining structure-function relationships across individuals or populations, as such variability is not random but regionally structured and functionally meaningful. For example, \cite{Stoecklein2020} showed that regions associated with higher-order cognitive functions (e.g., association cortices) exhibit greater inter-individual variability, whereas unimodal regions (e.g., visual and motor cortices) are more conserved across individuals. In addition, the study from \cite{Sun2022} identified a spatial alignment between structural and functional connectome variability, characterized by greater variability in heteromodal association regions and reduced variability in primary regions. The study revealed that functional variability can be robustly predicted by the structural variability pattern, following a hierarchical primary-to-heteromodal axis \citep{Sun2022, Huntenburg2018}, with predictions being more accurate in the primary regions and less so in the heteromodal regions. This spatial heterogeneity suggests that increases or decreases in variability may reflect differences in developmental constraints, functional specialization, and flexibility, thereby providing important insights into how structural and functional networks are coupled. 

In this paper, we define the structural-functional module consistency (SFMC) as a probabilistic measurement characterizing the alignment of the modular structure of SC and FC at the group level, meanwhile taking into account the inter-individual variability. The value of SFMC is calculated over a group of subjects within a specific age range during the infant brain development. Specifically, for individual-level SC or FC in that age range, we use the maximization of modularity quality function to estimate deterministic modular structure quantified by individual-level module labels with different modularity resolutions \citep{Bassett2013}. For group-level SC or FC networks, we model the estimated individual-level module labels across subjects using a Bayesian framework based on the categorical-Dirichlet conjugate pair \citep{Bian2026}, and define a label assignment probability vector (conceptualized as a stochastic module) that quantifies the likelihood that a given node is assigned to a specific module in either the structural or functional network. Within this framework, a brain region is no longer allocated within a deterministic module in a brain network, but is instead dynamically integrated or segregated into stochastic modules across subjects. Finally, SFMC is computed as the correlation between the label assignment probability vector of the group-level SC and that of the group-level FC for a given node. 

From a neuroscientific perspective, the correlation between the label assignment probability of a node in SC and that of the same node in FC reflects the degree of potential overlap between the SC and FC modules to which this node may belong. This provides a regionally specific measure of how structural organization constrains functional flexibility at the network level. The developmental trajectory of SFMC therefore captures age-related changes in this anatomical constraint, revealing how the alignment between structural and functional modular organization evolves during early brain development. Importantly, this framework enables the identification of regionally heterogeneous developmental patterns in structure-function relationships, particularly highlighting distinct trajectories between primary sensory regions and higher-order association cortices. Specifically, we observe differential patterns of structural-functional alignment, suggesting earlier stabilization of structure-function relationships in sensory systems and more prolonged reorganization in association cortices. Such module-level reconfiguration cannot be readily captured by conventional structure-function coupling approaches, which rely on nodal connectivity profiles and do not explicitly account for modular organization. By shifting the focus from connectivity similarity to module-level consistency, the proposed SFMC framework provides new insights into the developmental reorganization of brain specialization and integration, revealing how the balance between structural constraint and functional flexibility is progressively refined across early development.

\section{Materials and methods}
\subsection{Data acquisition and preprocessing}

The imaging scans of BCP were collected using 3T Siemens Prisma MRI scanners equipped with 32-channel head coils at the Center for Magnetic Resonance Research (CMRR) at the University of Minnesota and the Biomedical Research Imaging Center (BRIC) at the University of North Carolina at Chapel Hill \citep{Howell2019}. Both rs-fMRI and dMRI were collected using single-shot EPI sequences. Since BCP is a longitudinal dataset, subjects may have more than one scans across different age periods, and anterior-to-posterior (AP) and posterior-to-anterior (PA) scans were employed to correct distortions caused by EPI sequence. In BCP, the imaging protocol was designed to be largely consistent across participants to facilitate comparisons of brain development across ages. The same MRI scanner, head coil, and core acquisition protocols were used for both infants and older children, including similar field of view (FOV) and sequence configurations for rs-fMRI and dMRI.

Prior to scanning, all participants removed metal objects. Preschool aged children changed into facility-provided scrubs, and infants were swaddled in an MRI compatible blanket. For sleeping scans, children acclimated in a dimly lit preparation room, where caregivers followed typical bedtime routines (e.g., diaper change, feeding, reading). Earplugs were inserted and secured with tape before entry to the scanner room. For awake scans, normal lighting was maintained. Children selected a movie to view during the scan, and the MRI technologist confirmed safety compliance for both the child and caregiver.

\begin{figure*}[!ht]
\centering
\includegraphics[width=0.93\linewidth]{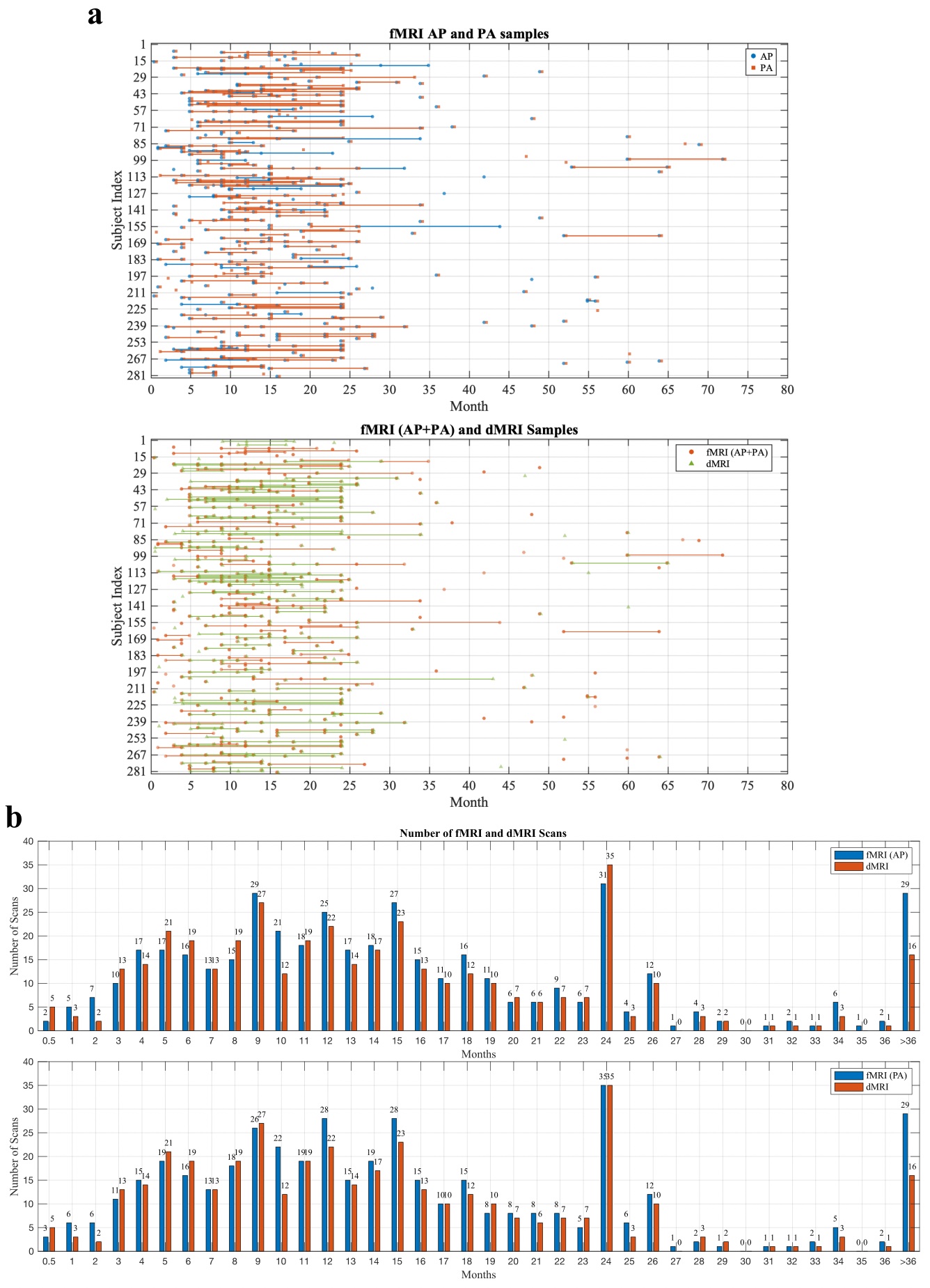}
\caption{\textbf{Samples acquired from fMRI and dMRI imaging}. \textbf{a} Longitudinal scans were acquired at different ages (in months). The fMRI data include unequal numbers of AP and PA scans, whereas the dMRI data contain equal numbers of AP and PA scans. \textbf{b} Comparison of the age distributions of fMRI and dMRI scans (AP: anterior-posterior; PA: posterior-anterior; LH: left hemisphere, RH: right hemisphere).}
\label{subject_samples}
\end{figure*}

There are 281 subjects involved in this study (Fig.\ref{subject_samples}). In the case of rs-fMRI, there are 433 scans of anterior-posterior (AP) and 438 scans of posterior-anterior (PA), contributed by 257 subjects.
The rs-fMRI protocol parameters include FOV of 208 mm$\times$208 mm, a voxel size of 2 mm isotropic, a flip angle of 52$^{\circ}$, an echo time (TE) of 37 ms, multi-band factor of 8, a repetition time (TR) of 800 ms, and time course of 381 frames. The rs-fMRI preprocessing \citep*{Zhang2019} includes head motion correction, distortion correction, anatomical registration, one-step resampling, and denoising \citep*{Heo2022}. 

For dMRI, there are 392 AP and 392 PA scans (each subject has both AP and PA) from 214 subjects, the FOV is 210mm$\times$210mm, voxel size is 1.5mm isotropic, flip angle is 78 excite/160 refocus, TE is 88.6ms, TR is 2640ms, and multi-band factor is 5. The dMRI preprocessing \citep{Liu2024} includes bias correction \citep{Andersson2003, Raffelt2017, Kellner2016}, distortion correction \citep{Andersson2016}, denoising \citep{Cordero-Grande2019}, rotation and correction of the B-matrix \citep{Andersson2016a}, fiber orientation distribution function generation and normalization \citep{Dhollander2021}, and probabilistic fiber tracking \citep{Tournier2010}. The dMRI data have $144$ directions  placed on  $6$ shells ($b$-values = $500$, $1000$, $1500$, $2000$, $2500$, and $3000$\,s/mm$^{2}$), along with  $6$ $b_0$ volumes. To reliably tract streamlines, we exploited the anatomically constrained tractography (ACT) framework implemented in \textsc{MR}trix3 \citep{Tournier2019}. The necessary GM/WM interface was obtained by tissue segmentation using our proposed sMRI segmentation tools specifically designed for infants \citep{Liu2024b}. Starting from seeds locating on the GM/WM interface, we tracked streamlines using the MSMT-CSD algorithms \citep{Tournier2010}. Eventually, SIFT2 \citep{Smith2018} was employed to mitigate the false positive issue.

\begin{table}[!ht]
\caption{Age ranges with different numbers of scans for fMRI and dMRI.}
\centering
\begin{tabular}{|c|c|c|c|c|c|c|c|c|c|r|l|} \hline
 Age ranges (month) & 0-5 & 3-8 & 6-11 & 9-14 & 12-17 & 15-23 & 18-29 & 24-36 & $>$36  \\ \hline
fMRI Scans (AP) & 58 & 88 & 112 & 128 & 113 & 107 & 108 & 67 & 29  \\ \hline
fMRI Scans (PA) & 60 & 92 & 114 & 129 & 115 & 105 & 109 & 68 & 29  \\ \hline
dMRI Scans (AP) & 58 & 99 & 109 & 111 & 99 & 95 & 102 & 60 & 16  \\ \hline
dMRI Scans (PA) & 58 & 99 & 109 & 111 & 99 & 95 & 102 & 60 & 16  \\ \hline
\end{tabular}
\label{agewindows}
\end{table}

\subsection{Construction of SC and FC}
The FC (with 100, 200, 300, and 400 ROIs) was constructed by calculating the correlation matrix of the extracted BOLD signals using the Schaefer's atlas \citep*{Schaefer2018}. In this study, negative FC values were set to zero to maintain interpretability, analytical stability, and compatibility with standard graph-theoretical methods, as well as to reduce the influence of potentially artifactual anti-correlations introduced by preprocessing. The SC (with 100, 200, 300, and 400 ROIs) was constructed using MRtrix3 (tck2connectome). Streamlines were first processed using SIFT2, which assigns a biologically informed weight to each streamline, proportional to the estimated fiber cross-sectional area. Edge weights were then computed as the sum of SIFT2-weighted streamlines connecting each pair of regions. In addition, to account for differences in ROI size, each streamline contribution was further scaled by the inverse of the volumes of the two connected nodes (using the -scale\_invnodevol option). Thus, the resulting connectivity represents a volume-normalized, SIFT2-weighted estimate of SC between regions.

\begin{figure*}[!ht]
\centering
\includegraphics[width=0.85\linewidth]{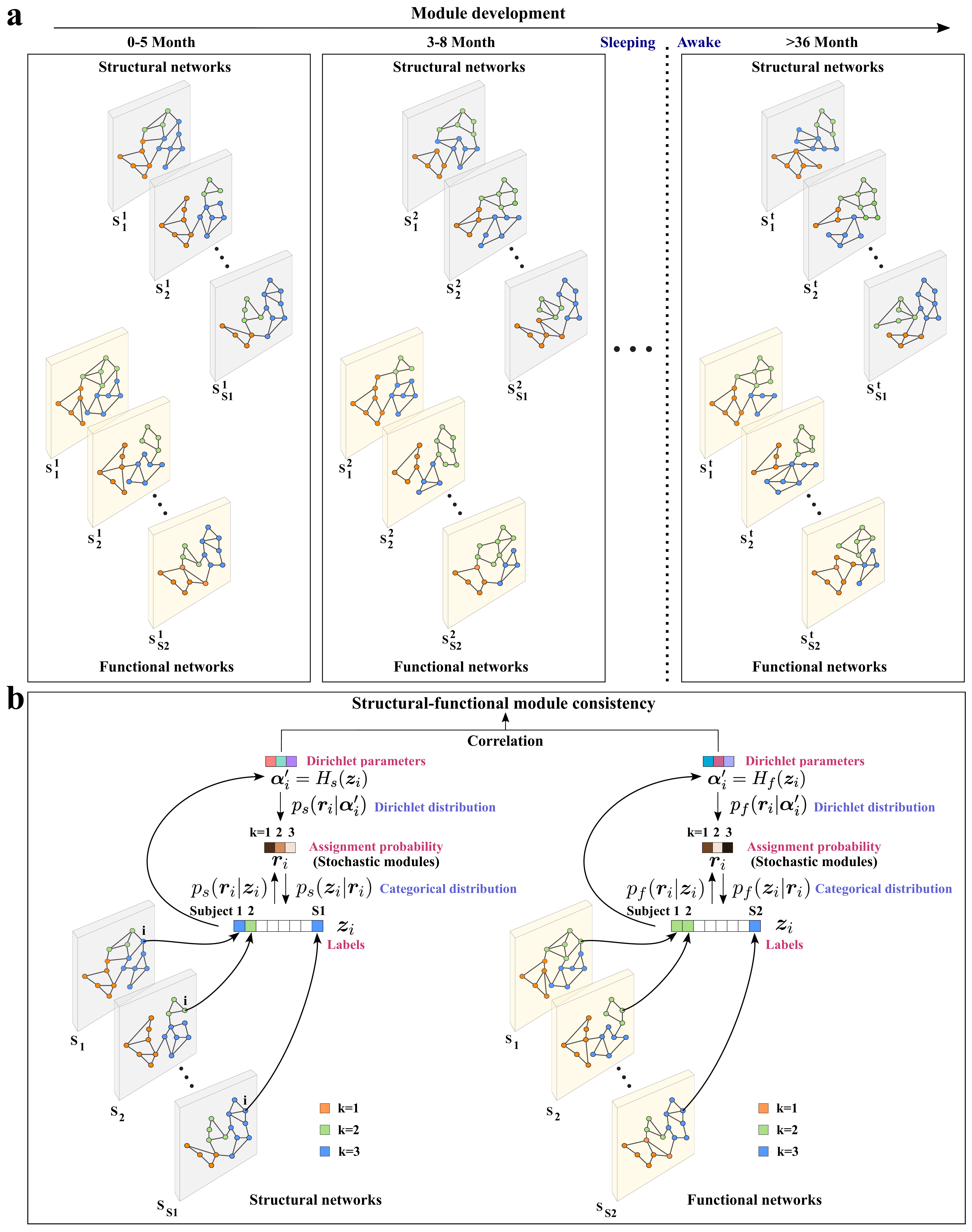}
\caption{\textbf{Overview of the framework.} \textbf{a} The module development for structural and functional networks. The participants were grouped into different age ranges, with those under 36 months being scanned while asleep, and those over 36 months being scanned while awake. The nodes are colored differently to indicate various module labels. \textbf{b}  Structural-functional module consistency (SFMC). The structural networks are constructed by estimating the fiber tracts between brain regions of interest (ROIs) for each subject (from subject 1 to $S1$ for a specific age range). The functional networks of each subject (from subject 1 to $S2$) are constructed by computing the Pearson's correlation from the BOLD time series extracted from brain ROIs. A vector of labels $\bm{z}_{i}$ of node $i$ represents the module labels across subjects drawn from the categorical distribution $p_{s}(\bm{z}_{i}\vert \bm{r}_{i})$ and $p_{f}(\bm{z}_{i}\vert \bm{r}_{i})$ for structural and functional networks respectively, conditional on a latent vector of assignment probability $\bm{r}_{i}$. Given $\bm{z}_{i}$, one can estimate a vector of assignment probability $\bm{r}_{i}$ by drawing the samples from the posterior distribution $p_{s}( \bm{r}_{i}\vert\bm{z}_{i})$ and $p_{f}( \bm{r}_{i}\vert\bm{z}_{i})$ for structural and  functional networks respectively, which is equivalent to drawing the samples $\bm{r}_{i}$ from the Dirichlet distribution $p_{s}( \bm{r}_{i}\vert\bm{\alpha}_{i}')$ and $p_{f}( \bm{r}_{i}\vert\bm{\alpha}_{i}')$. Here, the vector of Dirichlet parameters $\bm{\alpha}_{i}'$ is a function of the observation $\bm{z}_{i}$. The SFMC is calculated as the correlation between the vector of Dirichlet parameters of structural and functional networks.}
\label{SC_FC_network_Dirichlet_allocation}
\end{figure*}

\subsection{Data grouping}
The fMRI and dMRI datasets were longitudinally partitioned into different age groups using sliding windows with a six-month length during early development (0-17 months; e.g., 0-5, 3-8 months), while larger age window lengths were adopted for later developmental stages as shown in Table \ref{agewindows}, with varying numbers of subjects within each age range for the measurement of both SC and FC networks (see alternative age grouping in SI.8 and SI.9). Early brain development is highly nonlinear and rapid, and developmental changes can occur over short time scales. A six-month window provides a reasonable balance between temporal resolution and statistical stability, allowing us to capture developmental trends while maintaining sufficient sample sizes for reliable estimation of network measures within each range. The windows were shifted by three months, creating overlapping bins. This sliding-window strategy reduces discontinuities that may arise from arbitrary age boundaries and enables smoother characterization of developmental trajectories across age. In later infancy, broader age bins (e.g., 15-23 and 18-29 months) were employed to ensure sufficient sample sizes within each group, while also reflecting the slower pace of developmental change during this period. In contrast, too small windows would substantially reduce the sample size per bin, particularly at the youngest and oldest ages. 
\subsection{Overview of our framework}
The segregation and integration of networks undergo rapid changes during infancy in both brain structure and function (Fig.\ref{SC_FC_network_Dirichlet_allocation}a). We aim to investigate the relationship between modules in the SC network and FC network and how such relationship evolves with age. For that purpose, we define the SFMC as a probabilistic measurement of the modular structure consistency between structural and functional networks in a group of subjects.

The SC or FC constructed from each scan of a particular subject can be viewed as an observation of the underlying group representative network, which reflects the network characteristics of a population group. We first detect the modules from the individual-level SC or FC network based on \textit{modularity} \citep{Newman2004} separately for both SC and FC networks (Fig.\ref{SC_FC_network_Dirichlet_allocation}b). We then use Bayesian modeling at the group level to estimate the stochastic modules of the group representative network, while also accounting for inter-individual variability within each age range. 
We characterize the stochastic modules by an assignment probability for each brain region within the network over a group of subjects. This probability determines the likelihood of each brain region being associated with a module. The SFMC is then evaluated by the similarity of the distributions of stochastic modules of SC and FC.

\subsection{Individual-level modelling}

The modules of each individual SC or FC which are deterministic for each subject are detected by maximizing a modularity quality function \citep{Newman2006, Delvenne2010, Lambiotte2008} defined as 
\begin{equation}
Q=\frac{1}{2m}\sum_{i,j}[A_{ij}-\gamma\frac{k_{i}k_{j}}{2m}]\delta(g_{i},g_{j}),
\end{equation}
where $A_{ij}$ is the weight of connectivity between node $i$ and $j$, and $k_{i}=\sum_{j}A_{ij}$ is the strength of node $i$, $m=\frac{1}{2}\sum_{i,j}A_{ij}$ is the sum of the edge strength of the network. The delta function $\delta(g_{i},g_{j})=1$, if node $i$ and $j$ are in the same module ($g_{i}=g_{j}$), and 0 otherwise. The number of modules is controlled by the resolution parameter $\gamma$. Larger value of $\gamma$ corresponds to larger number of modules. Varying values of $\gamma$ result in coarse-to-fine network partitions. In this work, the maximization of $Q$ is implemented by a MATLAB function \textit{modularity\_und.m} in Brain Connectivity Toolbox \citep{Rubinov2010}. 

Since modules are detected at the individual level, the module labels of different subjects may have the label-switching phenomenon. For example, to represent a modular structure, a latent label vector $\lbrace 1,1,2,2,2,3,3 \rbrace$ of a network has the same meaning of $\lbrace 2,2,3,3,3,1,1 \rbrace$, with only module labels switched. To align module indices or labels across difference subjects, we adopted a relabelling algorithm \citep{Nobile2007,Wyse2012} based on minimization of label vector distance and square assignment algorithm \citep{Carpaneto1980} to obtain a switched module labels cross different subjects, so that the labels among subjects are maximumly consistent with each other.

We identified network modules separately for the left (LH) and right (RH) hemispheres. Functional networks often comprise both predominantly bilateral and unilateral modules, whereas structural networks are largely unilateral, consistent with \cite{Puxeddu2022}. For example, the left and right visual regions belong to a single module in the functional network but are assigned to distinct modules in the structural network. Consequently, evaluating structure-function modular consistency at the whole-brain level may introduce bias and result in misaligned modules within either left or right hemisphere.

\subsection{Group-level modelling}

Next, we will detail the use of conjugate Bayesian pairs to model the module labels of networks derived from individual-level analyses within each age range (or subject group, see in Fig.\ref{SC_FC_network_Dirichlet_allocation}b). The stochastic modules are characterized by a vector of assignment probabilities inferred by sampling from the posterior distribution. In Bayesian inference, the posterior density is obtained by combining the prior and likelihood. When both the prior and posterior belong to the same distribution family, the prior is called a conjugate for the likelihood, creating a prior-likelihood conjugate pair \citep{Bian2020}.


Assume there are $S$ subjects in a specific age range for an image modality (functional, or structural). We define a vector $\bm{z}_{i}=(z_{i1},\cdots,z_{is},\cdots,z_{iS})$ containing the module labels of node $i$ over all of the subjects within an age range and model it by a categorical-Dirichlet conjugate pair. Here, $i$ is the index of node and $s$ is the index of subject. We define a vector $\bm{r}_{i}=(r_{i1},\cdots,r_{ik},\cdots,r_{iK})$ containing the label assignment probabilities (LAP), where $k$ represents the label of module. Here, $\sum_{k=1}^{K} r_{ik}=1$, and $K$ is the maximumly possible label of module in the whole network in the group. 

Then the labels of each node $z_{is}\sim \mbox {Categorical}(1;\bm{r}_{i})$ and the likelihood function can be written as
\begin{equation}
p(z_{is}\vert\bm{r}_{i},K)=\prod_{k=1}^{K}r_{k}^{I_{k}(z_{is})}, \mbox{where\ }  I_{k}(z_{is})=
\begin{cases}
1, \ \mbox{if}\ z_{is}=k\\
0, \ \mbox{if}\ z_{is}\neq k\\
\end{cases}.
\end{equation}
Consider a prior of Dirichlet distribution $\bm{r}_{i}\sim \mbox{Dirichlet}(\bm{\alpha})$
\begin{equation}
p(\bm{r}_{i}\vert K)=N(\bm{\alpha})\prod_{k=1}^{K}r_{k}^{\alpha_{k}-1}, 
\end{equation}
where $\bm{\alpha}=(\alpha_{1},\cdots,\alpha_{k},\cdots,\alpha_{K})$, and the normalization factor $N(\bm{\alpha})=\frac{\Gamma(\sum_{k=1}^{K}\alpha_{k})}{\prod_{k=1}^{K}\Gamma(\alpha_{k})}$. In this paper, we set $\alpha_{k}=1$.
Then the posterior can be formulated as
\begin{eqnarray}
p(\bm{r}_{i}\vert \bm{z}_{i},K)&\propto& \prod_{s=1}^{S} p({z}_{is}\vert \bm{r}_{i},K) \times p(\bm{r}_{i},K)\\
&=&\prod_{s=1}^{S}\prod_{k=1}^{K}r_{k}^{I_{k}(z_{is})}\times N(\bm{\alpha})\prod_{k=1}^{K}r_{k}^{\alpha_{k}-1}\\
&=& N(\bm{\alpha})\prod_{k=1}^{K}r_{k}^{\sum_{s=1}^{S}I_{k}(z_{is})+\alpha_{k}-1}\\
&=&\frac{N(\bm{\alpha})}{N(\bm{\alpha}_{i}')}N(\bm{\alpha}_{i}')\prod_{k=1}^{K}r_{k}^{\alpha_{k}'-1},
\end{eqnarray}
where $\bm{\alpha}_{i}'=(\alpha_{1}',\cdots,\alpha_{k}',\cdots,\alpha_{K}')$, $\alpha_{k}'=\sum_{s=1}^{S}I_{k}(z_{is})+\alpha_{k}$, and $N(\bm{\alpha}_{i}')=\frac{\Gamma(\sum_{k=1}^{K}\alpha_{k}')}{\prod_{k=1}^{K}\Gamma(\alpha_{k}')}$.
Thus, the posterior distribution $p(\bm{r}_{i}\vert \bm{z}_{i},K)$ follows a Dirichlet distribution, $\mbox{Dirichlet}(\bm{\alpha}_{i}')$.
The characteristics of the distribution $p(\bm{r}_{i}\vert \bm{z}_{i},K)$ are governed by the parameter $\bm{\alpha}_{i}'$ which is a function $H$ of the observations $\bm{z}_{i}$. This parameter is expressed as $\bm{\alpha}_{i}'=H_{s}(\bm{z}_{i})$ for SC and $\bm{\alpha}_{i}'=H_{f}(\bm{z}_{i})$ for FC. 

\subsection{Evaluating SFMC}
The individual SC or FC within the same age range can be seen as samples from a group-level probability distribution. Intuitively, the consistency between structural and functional modules can be assessed by comparing the similarity of these two probability distributions of SC and FC in relation to the stochastic modules. We represent the distribution of the stochastic modules for a specific node $i$ as a Dirichlet distribution, denoted as $p_{s}(\bm{r}_{i}\vert \bm{z}_{i})=p_{s}(\bm{r}_{i}\vert \bm{\alpha}_{i}')$ and $p_{f}(\bm{r}_{i}\vert \bm{z}_{i})=p_{f}(\bm{r}_{i}\vert \bm{\alpha}_{i}')$ for structural and functional networks, respectively. The SFMC for a specific modularity resolution $\gamma$ is then defined as $c_{i}=corr(H_{s}(\bm{z}_{i}),H_{f}(\bm{z}_{i}))$. However, fixing an arbitrary $\gamma$ produces a single deterministic partition for both SC and FC and may therefore introduce bias into the resulting SFMC estimates. In this study, we average SFMC across a range of modularity resolution parameters ($\gamma$) as part of our methodological framework, termed the averaged SFMC $\overline{c}_{i}$, to reduce dependence on any single $\gamma$ value and integrate information across multiple scales. This is equivalent to examining the  behavior of networks from multiple perspectives, ranging from coarse to fine resolutions, and summarizing it as an average trend \citep{Bassett2013}. Specifically, the individual-level modularity is assessed across different modularity resolutions 
$\gamma$ ranging from 1.01 to 1.4 in increments of 0.01 in the current work, which allows us to capture community organization from relatively coarse to moderately fine resolutions while maintaining stable and interpretable modular structures. The lower bound of $\gamma$, set slightly above 1, prevents the detection of overly coarse partitions with too few communities (e.g., fewer than six), which may fail to capture the known large-scale organization of brain networks. While, increasing $\gamma$ gradually yields finer community partitions. However, excessively large $\gamma$ values tend to produce highly fragmented modules with limited interpretability. We found that $\gamma$ values above approximately 1.4 begin to produce overly fragmented community structures in our data. Therefore, the upper bound 1.4 was chosen to avoid excessive fragmentation into very small or noise-driven communities.

\subsection{Head motion examination}
To examine the relationship between head motion and age for an assessment of motion characteristics across infant age ranges while accounting for gender and imaging site, we performed an individual-level regression analysis using the ROI-wise temporal derivative of the rs-fMRI signal variance (DVARS). Specifically, We analyzed age-related changes in DVARS while accounting for possible confounding factors such as gender and imaging site.

A multiple regression model (DVARS $\sim$ Age $+$ Gender $+$ Site) is used to evaluate the age effect while adjusting for gender and site differences. Then, a residualized model in which DVARS was first regressed on gender and site to obtain motion residuals, followed by a separate regression of these residuals on age (DVARS $\sim$ Age) to isolate the age effect independent of confounding variables.

\section{Results}

\begin{figure*}[!ht]
\centering
\includegraphics[width=1\linewidth]{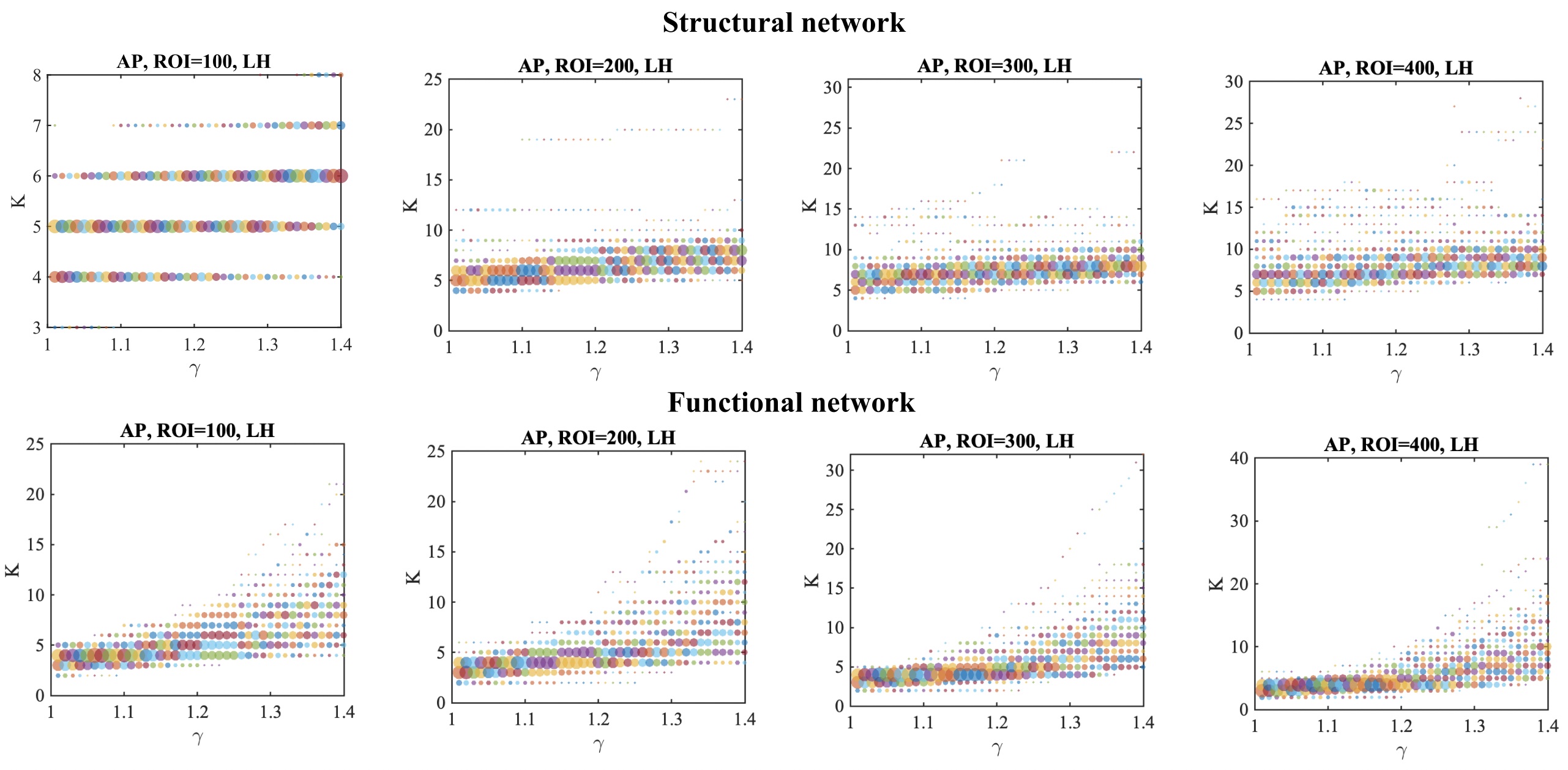}
\caption{\textbf{The number of modules with different modularity resolution $\gamma$}. The relationship between the number of modules $K$ and $\gamma$ with the size of the dots representing the number of subjects (a larger dot size indicates a greater number of subjects; AP: anterior-posterior; ROI: regions of interest; LH: left hemisphere).}
\label{module_number_SC_FC}
\end{figure*}

\begin{figure*}[!ht]
\centering
\includegraphics[width=0.9\linewidth]{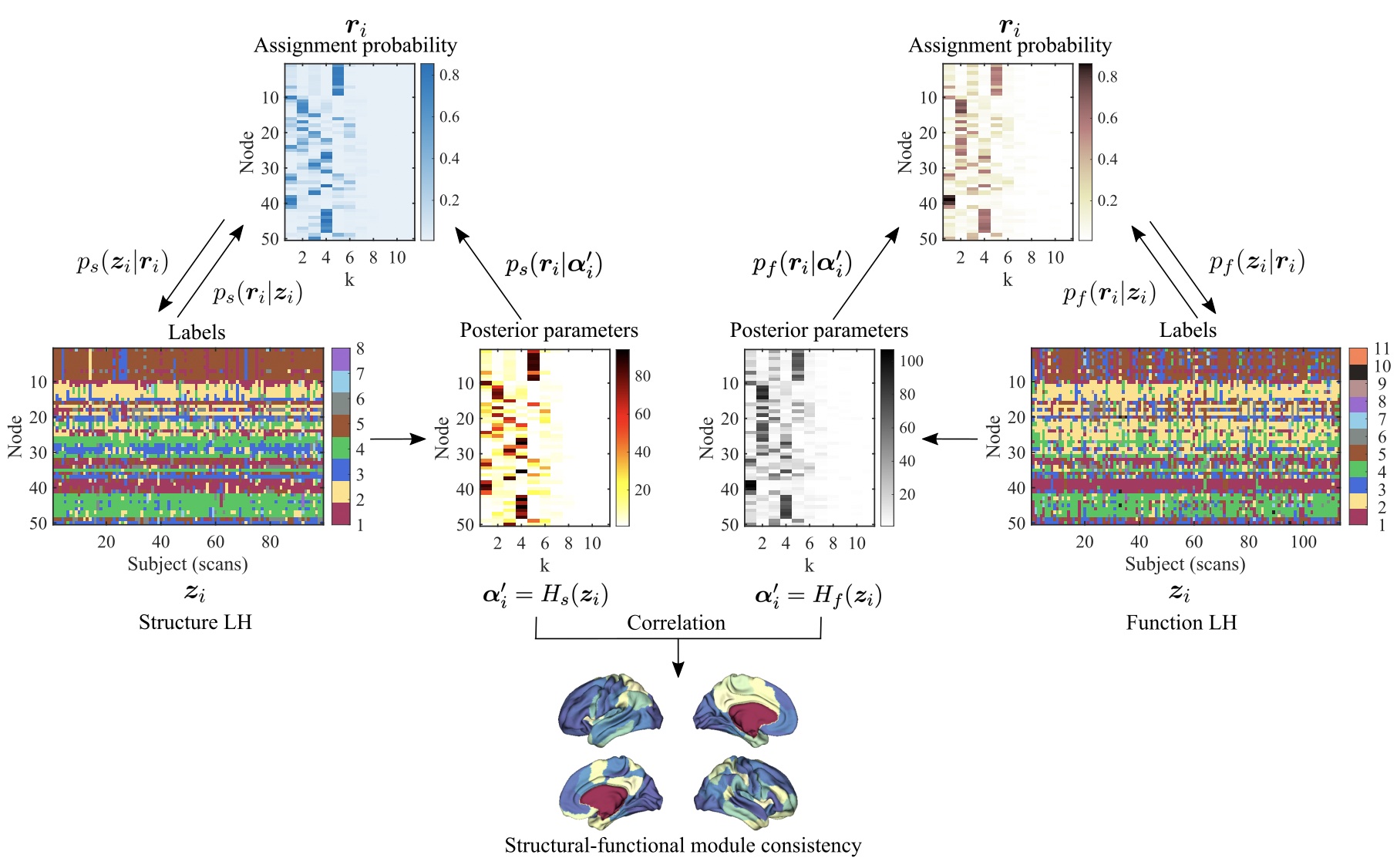}
\caption{\textbf{An example of structural-functional module consistency (SFMC) for the left hemisphere (LH) in the age range of 12-17 month (AP) with $\bm{\gamma}$=1.2}. In this age range, 99 scans were involved in constructing structural networks, while 113 scans were involved in constructing functional networks. The left hemisphere contains 50 nodes, making up 100 nodes in the entire brain. The different colors of the labels in $\bm{z}_{i}$ for a specific node $i$ indicate the module to which the node belongs. A vector of assignment probabilities $\bm{r}_{i}$ is determined by sampling from the posterior density using a Bayesian approach, with a vector of posterior parameters $\bm{\alpha}_{i}'$ dependent on the labels.}
\label{SC_FC_12-17}
\end{figure*}

\begin{figure*}[!ht]
\centering
\includegraphics[width=1\linewidth]{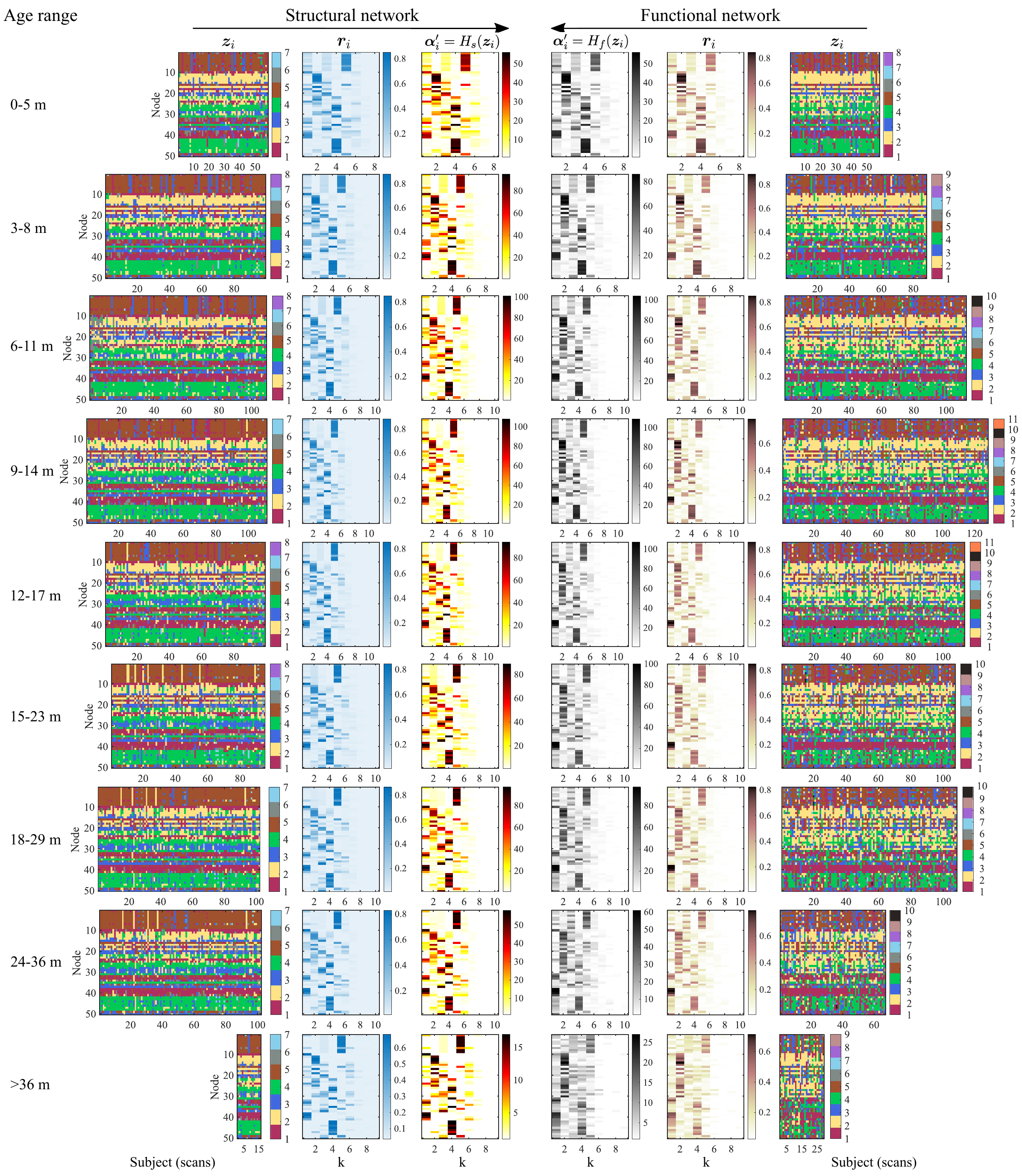}
\caption{\textbf{Modularity, assignment probability and Dirichlet parameters for different age ranges with $\bm{\gamma}$=1.2}. The scans of subjects are divided into different age ranges (from 0-5 month, 3-8 month, ..., and $>$ 36 month). The infants before 36 month were scanned under sleeping condition and those older than 36 month were scanned under awake condition. There are different numbers of scans of fMRI and dMRI in different age windows. The labels for each node in $\bm{z}_{i}$ are estimated by maximizing a modularity quality function. The assignment probability for each node $\bm{r}_{i}$ is inferred by drawing samples from the posterior density (a Dirichlet distribution). The vector of Dirichlet parameters is as a function of $\bm{z}_{i}$, that is $\bm{\alpha}_{i}'=H_{s}(\bm{z}_{i})$ for structural networks and $\bm{\alpha}_{i}'=H_{f}(\bm{z}_{i})$ for functional networks.}
\label{SC_FC_labels}
\end{figure*}

\begin{figure*}[!ht]
\centering
\includegraphics[width=0.9\linewidth]{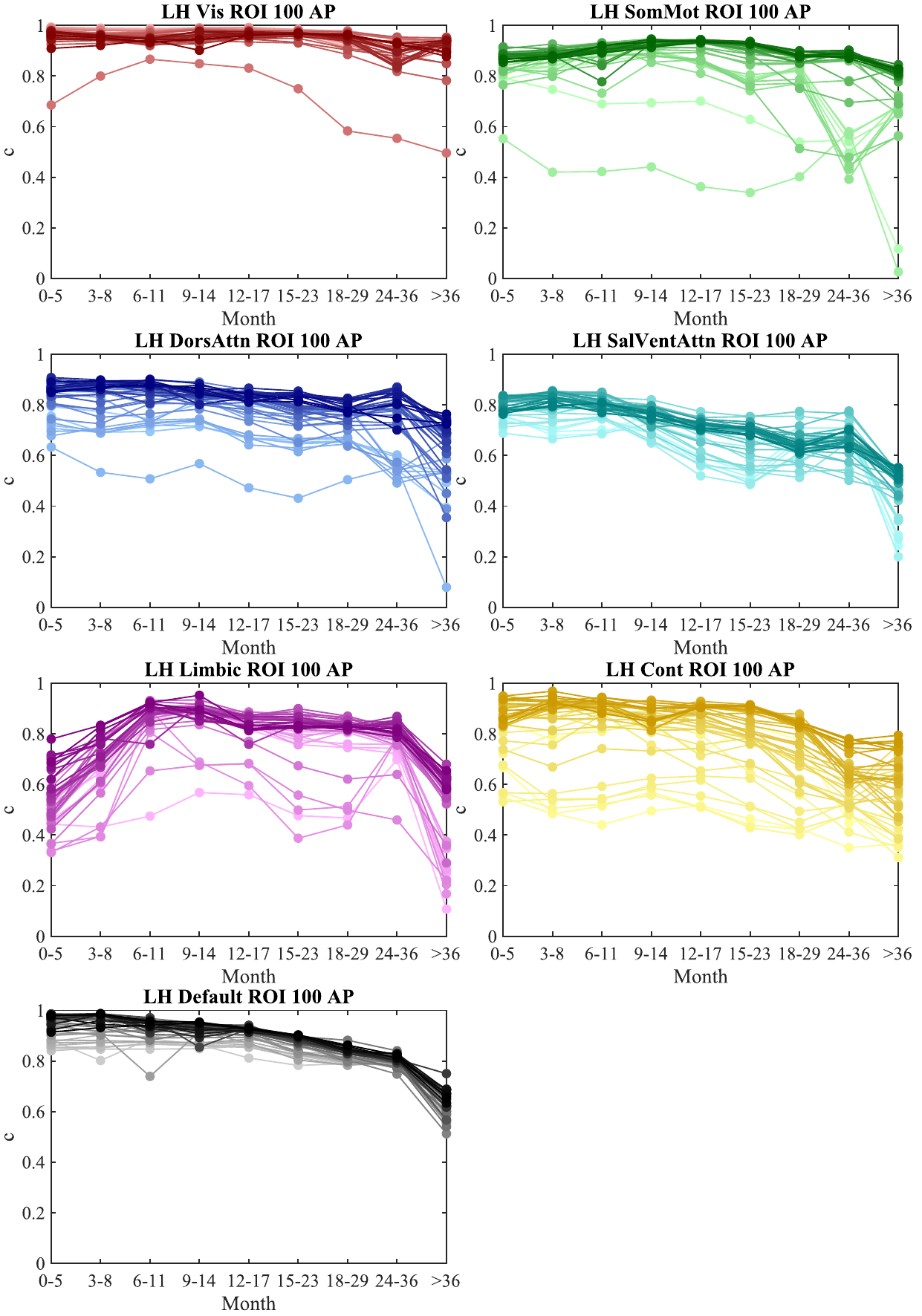}
\caption{\textbf{Variability of SFMC over $\gamma$ (AP)}. Each panel illustrates the c value of different brain networks (7-networks). Darker color represents larger value of $\gamma$ (from $\gamma$=1.01 to 1.4; AP: anterior-posterior; ROI: regions of interest; LH: left hemisphere)).}
\label{c_gamma}
\end{figure*}

\begin{figure*}[!ht]
\centering
\includegraphics[width=0.9\linewidth]{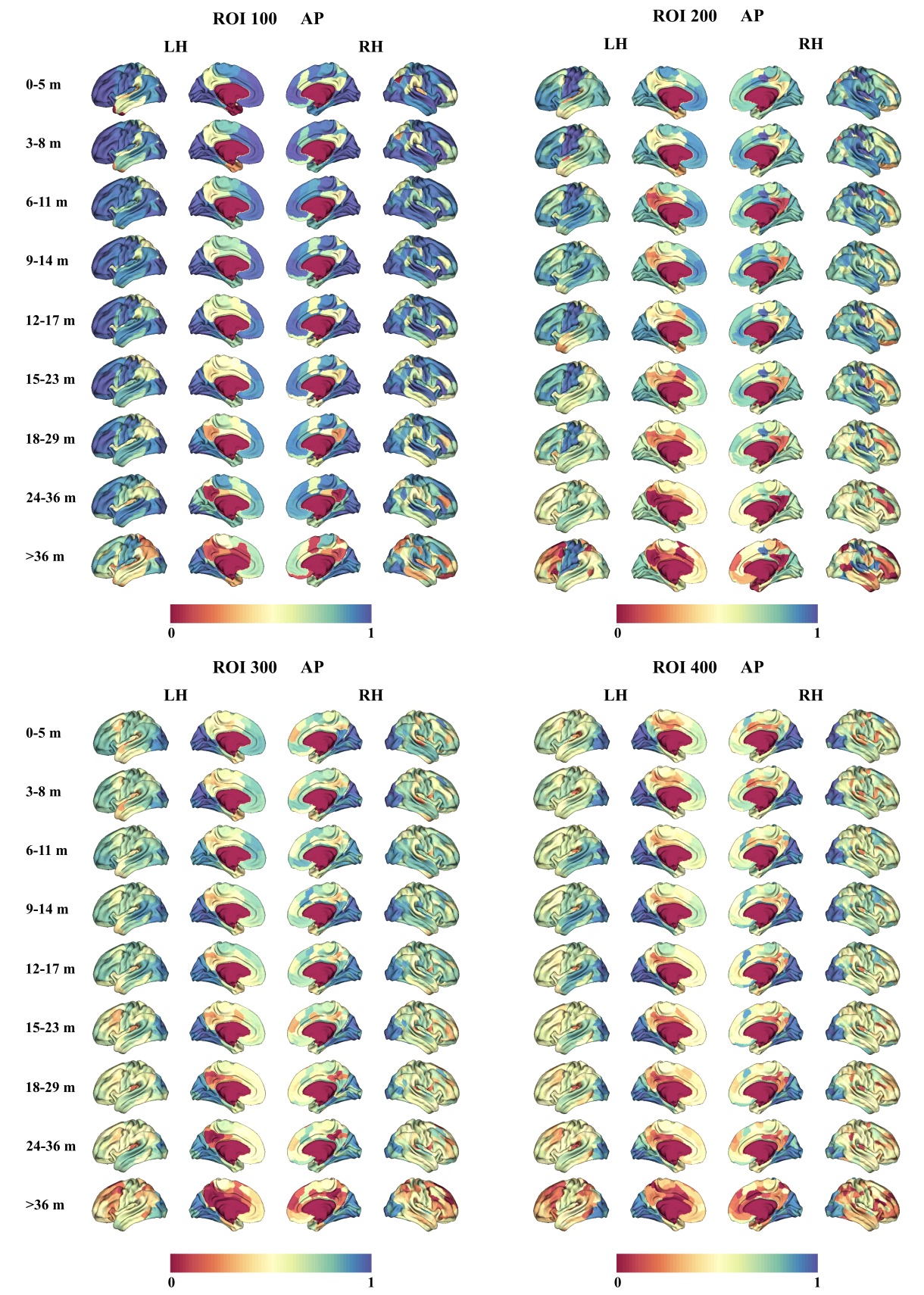}
\caption{\textbf{Averaged SFMC for different network scales from 100 to 400 ROIs}. The averaged SFMC of each ROI is calculated by averaging over the range of modularity resolutions $\gamma=1.01:0.01:1.4$ (AP: anterior-posterior; ROI: regions of interest; LH: left hemisphere, RH: right hemisphere).}
\label{Consistency_ap}
\end{figure*}

\begin{figure*}[!ht]
\centering
\includegraphics[width=0.9\linewidth]{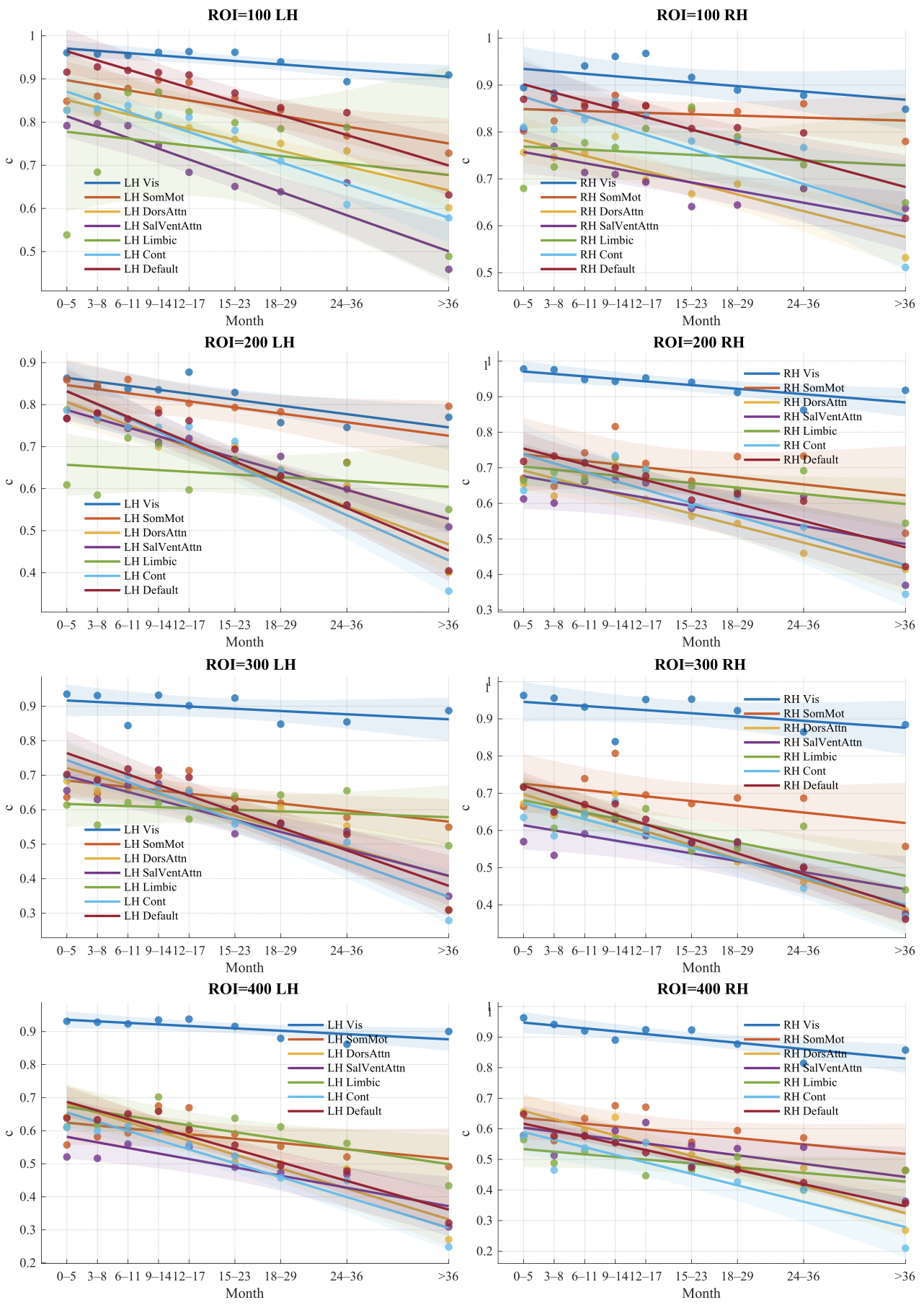}
\caption{\textbf{Mean value of the averaged SFMC within each of 7-networks based on stochastic modules (AP)}. Each panel illustrates the correlation analyses between mean value of the averaged SFMC and age (in months) for the left (LH) and right (RH) hemispheres for different scales (ROI = 100, 200, 300, and 400). Colored lines represent fitted regression curves for 7-networks. See the statistical analysis including the correlation (r), confidence interval (CI), and p-value in SI.1 (ROI: regions of interest; LH: left hemisphere, RH: right hemisphere).}
\label{Consistency_age_all_ap}
\end{figure*}

\subsection{Individual-level modelling}
The resolution parameter in the modularity model determines how many modules are identified in both structural and functional brain networks (Fig. \ref{module_number_SC_FC}); specifically, increasing $\gamma$ leads to a greater number of detected modules. Next, we demonstrate the results of estimation of labels using $\gamma=1.2$ of age range 12-17 (AP) month as an example (the left hemisphere of structural and functional networks (Structure LH and Funtion LH as shown in the most left and right side in Fig.\ref{SC_FC_12-17}). Each row ($\bm{z}_{i}$) represents the labels of a node across the subject scans in an age range, and each column contains the labels of the network from a specific scan. As we can see, the largest number of modules is 8 among all of the scans in that age range for SC, while the largest value is 11 for FC. Given a specific value of $\gamma$ ($\gamma=1.2$ for example), the modular structures of FC have larger variation across the subjects compared with that of the SC.

\subsection{Group-level modelling}
At the group level, the categorical-Dirichlet conjugate pair is used to model the module labels of each row vector in $\bm{z}_{i}$ estimated at the individual level. The module labels of each node across all of the subjects in a specific age range are regarded as the observations generated from $p_{s}(\bm{z}_{i}\vert \bm{r}_{i})$ conditional on a latent assignment probability $\bm{r}_{i}$ for structural networks. On the other hand, given $\bm{z}_{i}$, we can estimate the assignment probability which is inferred by sampling a posterior density $p_{s}(\bm{r}_{i} \vert \bm{z}_{i})$ which is formulated as a Dirichlet distribution $p_{s}(\bm{r}_{i} \vert \bm{\alpha}_{i}')$ with posterior parameters $\bm{\alpha}_{i}'$. Similarly, we also estimate the assignment probability from $p_{f}(\bm{r}_{i} \vert \bm{\alpha}_{i}')$ for FC. The SFMC is calculated as the correlation of $\bm{\alpha}_{i}'$ of the Dirichlet distributions of SC and FC. The estimated labels $\bm{z}_{i}$, the vectors of assignment probability $\bm{r}_{i}$, and the Dirichlet parameters $\bm{\alpha}_{i}'$ of SC and FC for different age ranges are referred to in Fig.\ref{SC_FC_labels}.

\subsection{SFMC across various modularity resolutions}
The modularity resolution parameter $\gamma$ affects the partition and number of modules. It is  interesting to examine how $\gamma$ impacts the group-level SFMC. Therefore, we further assess the SFMC across multiple modularity resolutions ($\gamma$) (Fig. \ref{c_gamma}). We found that increasing $\gamma$ generally leads to higher consistency within most sub-networks, except for the visual network. In addition, the SFMC decreases more dramatically for subjects of $>$ 36 month. It is noteworthy that the SFMC increases in the age ranges of 0-5, 3-8, and 6-11, before declining in the subsequent age ranges within the limbic network. Our method is robust to moderate variations in sample size; see the results for alternative age groupings in SI.8 and SI.9.

\subsection{Changes of SFMC}
The averaged SFMC $\overline{c}_{i}$ of the nodes is depicted in Fig.\ref{Consistency_ap} using BrainSpace \citep{VosdeWael2020} for the left (LH) and right (RH) hemisphere with different node scales from 100 to 400 ROIs. The $\overline{c}_{i}$ takes the value from 0 to 1 shown as a color bar, with the value closer to 1 indicating greater consistency. To depict $\overline{c}_{i}$ of the brain regions responsible for the specialized brain function, we calculate the mean value of $\overline{c}_{i}$ over the nodes within each of seven typical functional networks \citep{ThomasYeo2011} (see Fig. \ref{Consistency_age_all_ap}), including the visual network (Vis), the somatomotor network (SomMot), the dorsal attention network (DorsAttn), the salience or ventral attention network (SalVentAttn), the limbic network (Limbic), the control network (Cont), and the default mode network (Default). 
\subsection{Correlation analysis of SFMC}
According to Fig. \ref{Consistency_age_all_ap}, significant negative correlations were observed between age and mean $\overline{c}_{i}$ across all parcellation scales (ROI from 100 to 400) for the scan of AP (see SI.2 for the results of correlation analyses for PA), indicating that the coupling strength between structural and functional modules across all specialized brain networks progressively declined with age during infant brain development. The mean $\overline{c}_{i}$  of primary sensory regions is larger and decreases relatively slowly compared to the other areas of the brain. Advanced cognitive areas, including those related to attention, control, and default mode, show a prominent decrease with age (our method is robust to moderate variations in sample size; see the results for alternative age groupings in SI.8 and SI.9).

\subsection{SFMC differences across age groups associated with sleep and awake states}
We observed that SFMC decreases in the age range above 36 month compared to earlier age ranges across most subnetworks (Fig.~\ref{c_gamma}). Notably, these age ranges are associated with different vigilance states during data acquisition, with younger participants scanned during sleep and older participants during awake condition.

\begin{figure*}[!ht]
\centering
\includegraphics[width=1\linewidth]{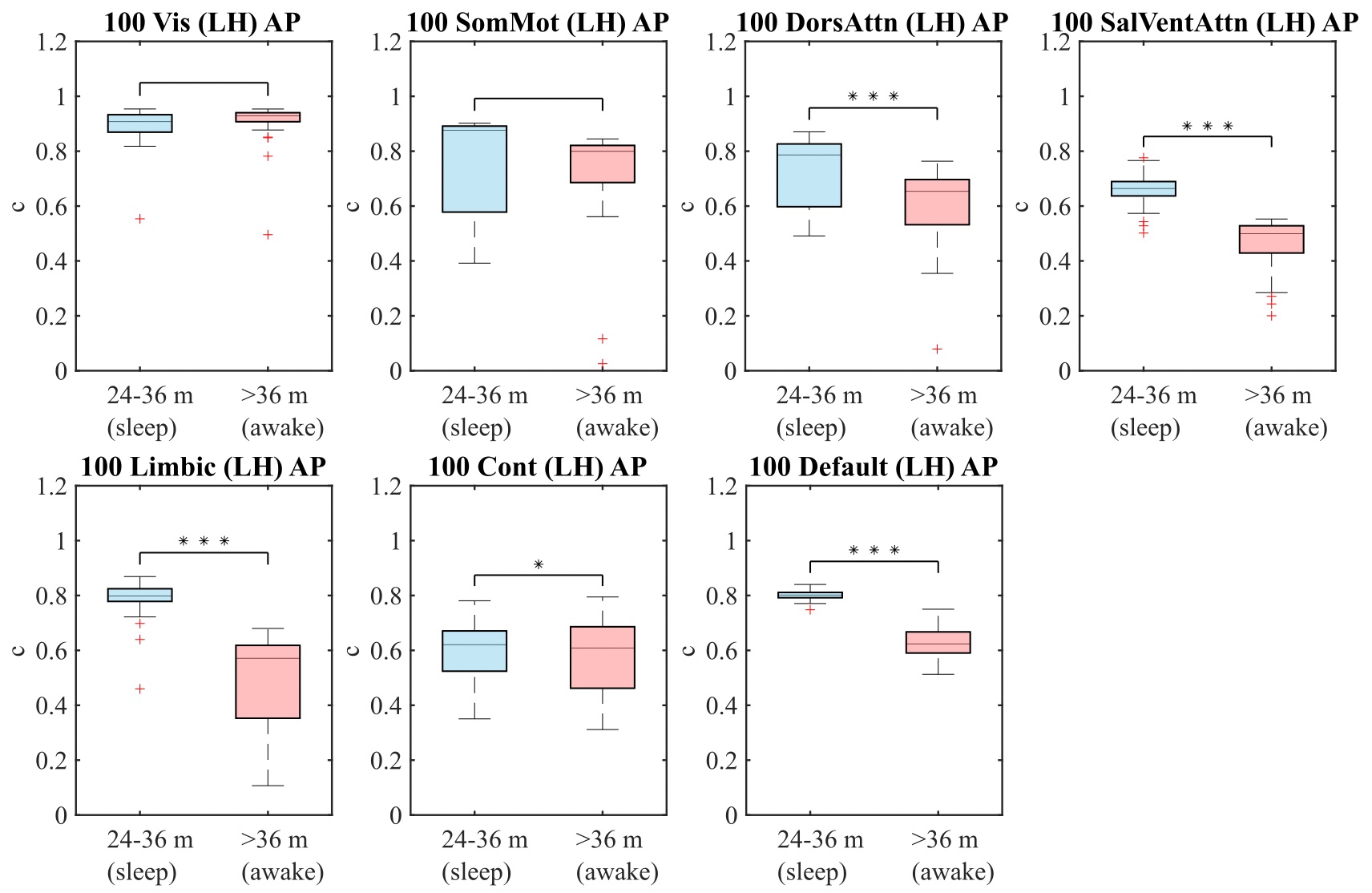}
\caption{\textbf{SFMC differences across age groups associated with sleep and awake conditions with ROI=100}. Wilcoxon signed-rank test with the asterisk indicates the statistical significance ($\ast$: $0.01\leq p<0.05$, $\ast$ $\ast$: $0.001\leq p<0.01$, and $\ast$ $\ast$ $\ast$: $p<0.001$; AP: anterior-posterior; LH: left hemisphere, RH: right hemisphere).}
\label{awake_sleep_test}
\end{figure*}

To further explore this observation, we compared SFMC values across different $\gamma$ between two groups: 24-36 month and above 36 month. Statistical analysis using the Wilcoxon signed-rank test revealed significantly lower SFMC in the latter group in several subnetworks, including the dorsal attention network (DorsAttn), salience/ventral attention network (SalVentAttn), limbic network (Limbic), and control network (Cont) (Fig.~\ref{awake_sleep_test}).

Importantly, we note that age and vigilance state are not independently controlled in the current dataset. Therefore, the observed differences should be interpreted as reflecting a joint effect of developmental stage and associated brain state, rather than a purely state-dependent effect. Disentangling the respective contributions of age and vigilance state would require data with overlapping age distributions across conditions or explicit covariate control, which is beyond the scope of the present study.

\subsection{Head motion analysis}

\begin{figure*}[!ht]
\centering
\includegraphics[width=0.9\linewidth]{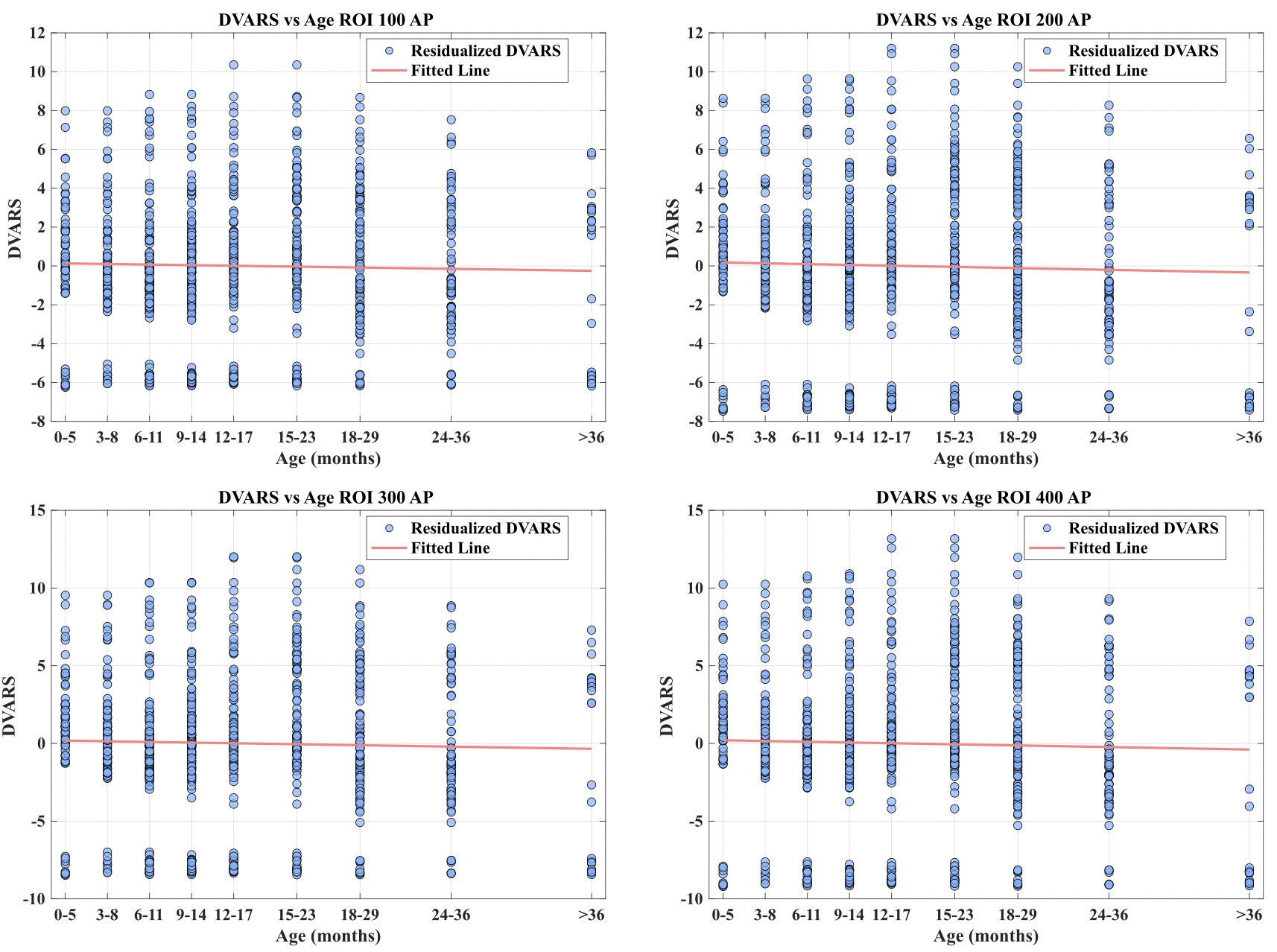}
\caption{\textbf{Regression of DVARS for different parcellation scales (ROI = 100, 200, 300, and 400, AP).}}
\label{DVARS}
\end{figure*}

A multiple linear regression model was performed to assess the influence of age, gender, and site on head motion, quantified by DVARS (Fig.\ref{DVARS}). The overall model was significant (F(3,806)=103, p$<$0.001), explaining approximately 27.8\% of the variance ($R^{2}$=0.278, adjusted $R^{2}$=0.275). Among the predictors, site showed a strong effect on DVARS ($\beta$=4.64, SE=0.28, t=16.79, p$<$ 0.001), whereas neither age ($\beta$=-0.011, p=0.45) nor gender ($\beta$=-0.44, p=0.09) contributed significantly to the model.

After residualizing DVARS by removing the effects of gender and site, a follow-up regression model examining the relationship between age and residualized DVARS revealed no significant age effect (F(1,808)=0.53, p=0.47, $R^{2}$=0.00065). These results indicate that head motion did not show a statistically significant systematic dependence on age after accounting for gender and site differences.

\subsection{Stratification analysis for confounders}
To account for potential confounders of gender and AP/PA encoding, the dataset was stratified by sex (female and male), and correlation analyses were performed separately for both AP and PA encoding directions. We found that neither gender nor encoding direction (AP/PA) altered the developmental trajectory of SFMC. The corresponding results are presented in SI (SI.3: female, AP; SI.4: female, PA; SI.5 male, AP; SI.6 male, PA).

\begin{figure*}[!ht]
\centering
\includegraphics[width=0.9\linewidth]{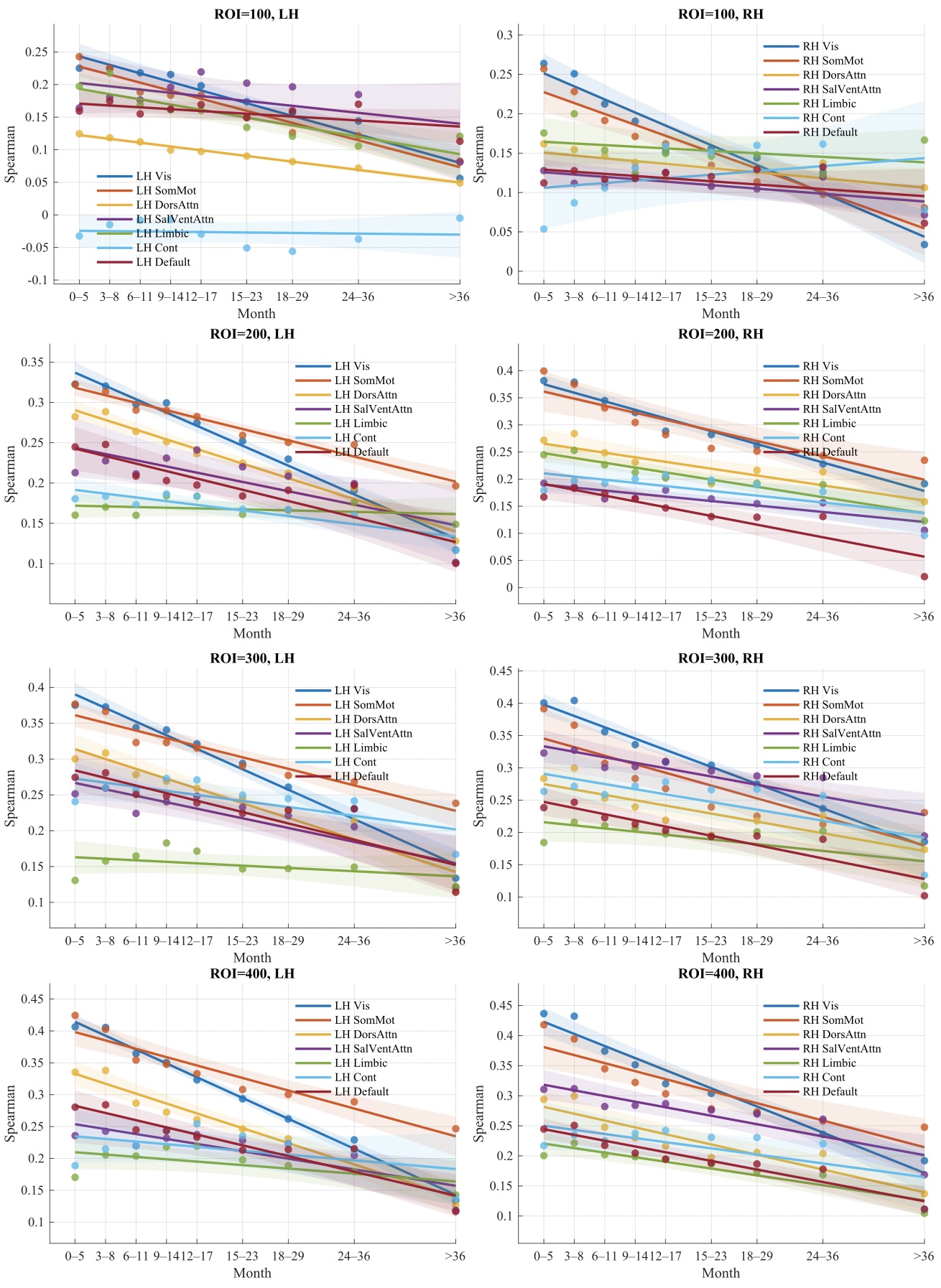}
\caption{\textbf{Structural-functional coupling based on Spearman correlation (AP)}. Each panel illustrates the correlation analyses between structural-functional coupling and age (in months) for the left (LH) and right (RH) hemispheres for different scales (ROI = 100, 200, 300, and 400). Colored lines represent fitted regression curves for 7-networks.}
\label{spearman_age_all_ap}
\end{figure*}

\begin{figure*}[!ht]
\centering
\includegraphics[width=0.9\linewidth]{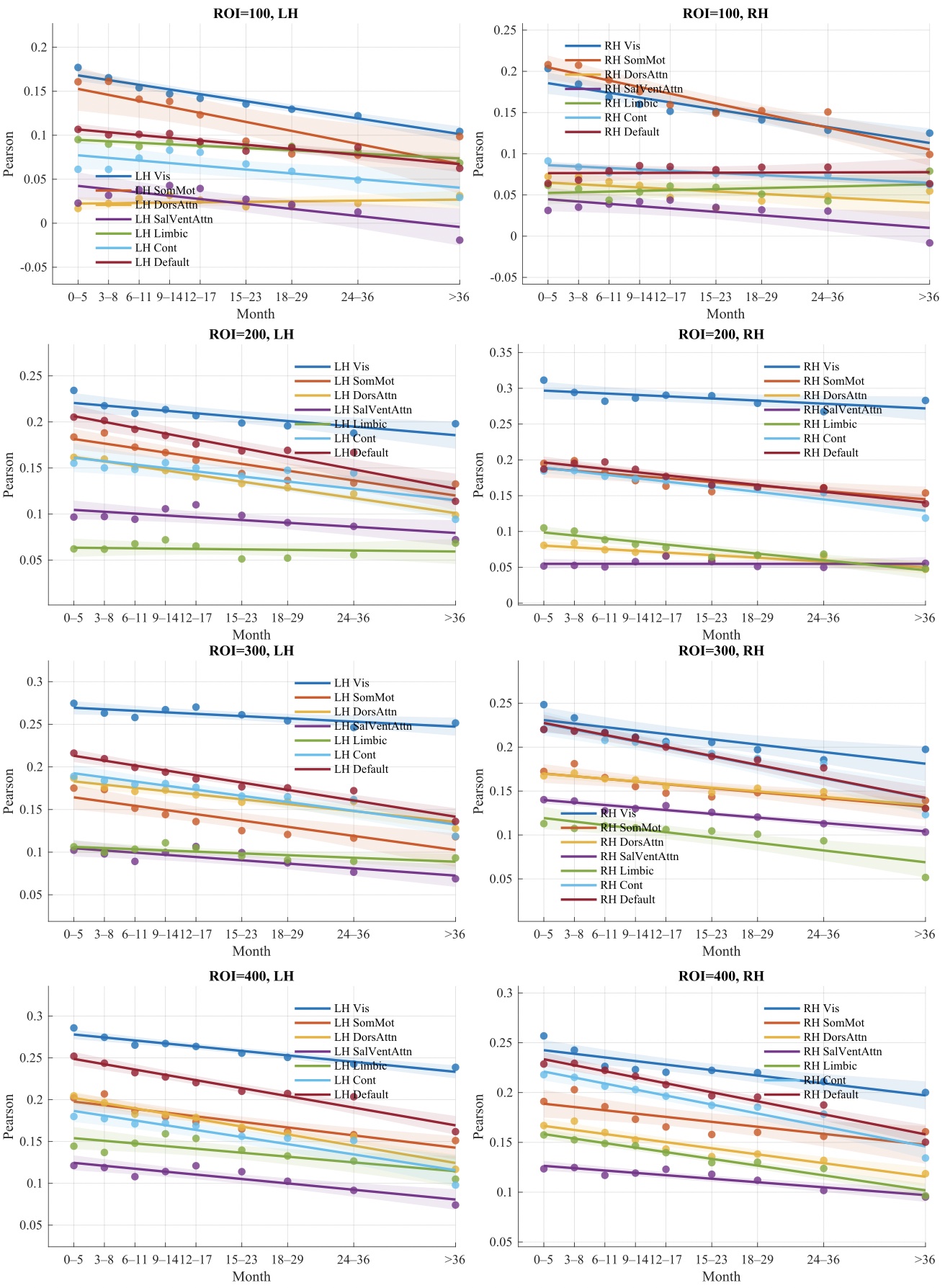}
\caption{\textbf{Structural-functional coupling based on Pearson's correlation (AP)}. Each panel illustrates the correlation analyses between structural-functional coupling and age (in months) for the left (LH) and right (RH) hemispheres for different scales (ROI = 100, 200, 300, and 400). Colored lines represent fitted regression curves for 7-networks.}
\label{pearson_age_all_ap}
\end{figure*}

\subsection{Comparison with traditional structural-functional coupling approach}
The standard approach for assessing structural-functional coupling involves using Spearman or Pearson correlation between the node profiles of SC and FC, as demonstrated in \citep{Baum2019}. Here, we compare our SFMC (Fig.\ref{Consistency_age_all_ap}) with these two conventional methods. A similar decreasing trend with age is observed using the structural-functional coupling based on Spearman correlation (Fig.\ref{spearman_age_all_ap}) and Pearson's correlation (Fig.\ref{pearson_age_all_ap}), but the decline is less steep than using SFMC, suggesting that while both conventional structural-functional coupling and SFMC weaken over time, SFMC shows a more pronounced developmental reorganization. In addition, it is difficult to distinct the developmental trajectory between the advanced cognitive areas and primary sensory regions using Spearman correlation.

\section{Discussion}

\subsection{Group-level versus individual-level dynamics}
In this work, each scan is treated as an independent sample within a given age range, and repeated measures from the same subject are not explicitly modeled. All scans are pooled and analyzed at the group level, so multiple scans from a single subject are considered separately. The analysis therefore emphasizes age-related trends across the cohort rather than individual longitudinal trajectories.

This framework accommodates unequal numbers of fMRI and dMRI scans and naturally handles missing data. Specifically, SC and FC observations within each age range are modeled as samples drawn from latent distributions representing group-level structural and functional networks, eliminating the need for matched scans across modalities or participants. As a result, the proposed SFMC captures group-level properties and developmental patterns of the cohort, rather than subject-specific changes.

\subsection{The difference in the physiological interpretation of SC and FC}
Although many works have explored the relationship between brain structure and function, an important concern is that SC and FC have distinct interpretations and topologies \citep{Puxeddu2022}. SC represents physical connections between brain regions, with weights reflecting white-matter fiber tracts that are always non-negative. Although their density can vary depending on factors like brain region dimensions (e.g., parcellation) or reconstruction methods, they are typically sparse. In contrast, FC reflects statistical relationships between the BOLD activity of different brain regions, leading to fully connected networks with both positive and negative weights. This indicates that SC and FC have distinct physical definitions, and thus, directly calculating the correlation between their profiles (such as in conventional structure-function coupling) may not accurately represent the true relationship between brain structure and function. Instead, our approach indirectly assesses the relationship between the modular structures of SC and FC by examining the correlation of parameters from the latent probability distributions, offering a more meaningful evaluation of the SFMC from a statistical standpoint.

\subsection{Inter-individual variability versus group average}
Many studies build group representative or group-level networks by calculating a group-averaged connectivity matrix, using FC as an example \citep{Robinson2015, Bian2021}, which overlooks the differences and variations in networks across individuals. The variations in SC and FC patterns result not only from anatomical differences between subjects and unpredictable fluctuations in latent cognitive states during resting periods, but also from non-neural physiological factors. These include head motion, cardiovascular and respiratory influences, as well as noise caused by hardware instability \citep*{Hutchison2013, Lurie2020}. 
The presence of noise can alter both SC and FC between pairs of brain regions, potentially leading to significant changes in the modular structure. Considering only a single metric, such as the group-averaged connectivity matrix, may lead to a biased estimation of the modular structure. Apart from ignoring information that is shared across individuals \citep{Lehmann2021}. Another issue is that the outlier of some individual SC or FC may result in biased estimation for calculating group averaged connectivity matrix. Modelling individual networks instead of the group-averaged network offers insights into the variability of both the observations and the modular structure across subjects. 
\subsection{SFMC in typical brain sub-networks}
According to our results, sub-network specific SFMC decreases with age, suggesting a progressive reduction in the relationship between brain structure and function during early development. The observed decrease in SFMC from infancy to early childhood may reflect the increasing functional flexibility of brain networks during development. Early brain organization is thought to be strongly constrained by structural connectivity \citep{Honey2009}, whereas functional interactions become progressively more dynamic and less tightly coupled to anatomical pathways as higher-order cognitive systems mature. This finding is consistent with previous reports showing that average structure-function coupling is negatively associated with age in early life, with pronounced decreases observed in primary sensory systems, particularly in auditory and somatomotor regions \citep{Tooley2025}. The age-related decoupling between structural and functional connectivity may reflect developmental changes in brain organization.

Additionally, the SFMC results exhibit regionally structured patterns that are functionally meaningful. The trajectories of averaged SFMC across nodes in the seven networks suggest that the primary sensory regions of the visual cortex exhibit relatively higher SFMC and a slower decline. This result is consistent with previous works that evolutionarily conserved primary sensory regions (e.g., visual and somatomotor cortices) show relatively strong structural-functional coupling, whereas rapidly expanded transmodal association regions (e.g., limbic and default mode networks) exhibit weaker structural-functional coupling \citep{Gu2021}. Higher SFMC in primary sensory regions, such as the visual cortex, likely reflects the early maturation and strong anatomical constraints of sensory systems. In contrast, association cortices involved in attention, cognitive control, and the default mode network demonstrate lower consistency, possibly reflecting their protracted developmental trajectories and increased functional flexibility. This hierarchical maturation pattern of sensory and association systems has been consistently documented in developmental neuroscience \citep{Gao2015}. Primary functional networks tend to mature earlier, showing synchronized activity already at birth, whereas higher-order networks undergo more extended postnatal periods of functional synchronization \citep{Gao2014}. Brain maturation generally progresses from primary sensory networks to networks associated with attention and self-awareness, and ultimately to those supporting higher-order executive functions.

Notably, SFMC within the somatomotor network, another component of the primary sensory system, shows a decreasing trend toward higher-order cognitive regions, although the decline is less pronounced than that observed in other networks. This difference may reflect the broader interactions of somatomotor regions with higher-order cortical areas, compared with the more specialized visual cortex. As a result, functional activity in the somatomotor network may be less tightly constrained by local structural pathways, which could contribute to different SFMC trajectories compared with those observed in the visual cortex. This is consistent with other reported studies that cortices at the sensorimotor pole of the sensorimotor-association axis exhibited substantial increases in connectivity during development \citep{Luo2024}. These regions became more strongly connected with other sensorimotor cortices and with middle-axis regions, including attention systems. In contrast to other sensorimotor areas, the primary visual cortex showed a decrease in functional connectivity strength across development, which may reflect the requirement for more segregated visual processing. Overall, these results suggest increasing functional integration of somatomotor regions and greater cross-system coherence. This supports the idea that somatomotor regions interact more broadly across networks, which could weaken SFMC.

Finally, we discuss the differences across age groups associated with sleep and awake states. Although age and vigilance state are not independently controlled in the current dataset, we speculate that the lower SFMC and faster developmental decline observed during wakefulness may reflect greater cognitive engagement and more dynamic integration of functional networks in the awake brain, whereas sleep is associated with more stable and structurally constrained network activity. This interpretation is supported by previous findings showing that although large-scale brain modules (e.g., the default mode network) remain present during sleep, functional integration is substantially reduced \citep{Boly2012, Tagliazucchi2013a}. Moreover, prior work has reported that temporal integration, reflecting the long-term memory of neural activity, progressively decreases from wakefulness to deep sleep, with pronounced effects in regions associated with the default mode and attention networks \citep{Tagliazucchi2013}.

\subsection{Correlation analysis}
In this study, correlation analysis was used to evaluate the relationship between SFMC and age because SFMC is calculated at the group level, resulting in a limited number of observations. Specifically, each age range yields a single summary value across subjects rather than individual-level measurements. With only a small number of group-level data points, the sample size is insufficient for regression modeling, which requires estimating multiple parameters. Correlation analysis, which tests the covariation between two variables, is more appropriate under such small-sample conditions when the relationship is approximately linear.

\subsection{limitations}
Some limitations in this study should be acknowledged. One limitation is that our method may disrupt the assessment of within-subject dynamics. Because scans are treated independently, any longitudinal trends within a single subject (e.g., how their SC-FC coupling changes over time) are not directly modeled. Therefore, the analysis loses information about individual developmental trajectories, focusing only on the group pattern.

In addition, network modules were identified separately for each hemisphere. While this approach helps mitigate the influence of hemispheric asymmetries, particularly in structural networks. It may limit the assessment of whole-brain convergence between structural and functional organization. Specifically, functional networks often exhibit bilaterally distributed modules, which may not be fully captured under hemisphere-specific partitioning. As a result, our framework may underestimate the degree of structure-function convergence at the global scale. Future studies are needed to develop approaches that can simultaneously account for both bilateral and unilateral modular organization, enabling a more comprehensive characterization of structural-functional consistency.

In this study, DVARS was derived from ROI-wise BOLD signals, which are computed solely from the fMRI data by averaging voxel signals within each brain region. Consequently, this approach may not accurately capture the actual motion of the infants. Moreover, variations in brain parcellation scale (from 100 to 400 ROIs) may further affect the accuracy of the ROI-wise DVARS estimation.

Another key limitation of this study is that age and vigilance state (sleep vs awake) are not independently controlled in the current dataset. In typical infant MRI acquisition protocols, younger participants are more likely to be scanned during sleep, whereas older participants are more likely to be awake, leading to a partial collinearity between developmental stage and brain state. As a result, the observed SFMC differences between groups may reflect a joint effect of age and vigilance state, rather than a purely state-dependent or development-specific effect. Future studies with balanced sampling of sleep and awake conditions within the same age range, or with explicit covariate control, will be necessary to disentangle the respective contributions of developmental and state-related factors.

In this paper, standardized acquisition strategy was intentionally adopted in the BCP to ensure comparability of imaging data across developmental stages and to support longitudinal and cross-sectional analyses of brain development \citep{Howell2019}. Maintaining a consistent acquisition protocol reduces variability introduced by scanner settings and is a common practice in large developmental neuroimaging studies. However, using the same imaging protocol setup across a wide age range may introduce some limitations. For example, differences in head size and brain morphology between younger infants and older children could potentially affect signal coverage or spatial resolution relative to brain size \citep{Raschle2012}.

Infants may awaken during sleeping scans, sometimes becoming fussy, and other times remaining calm, making it unreliable to assume they are asleep merely because they appear still and quiet. However, no physiological signals (heart rate, respiratory traces) were recorded in the protocol of BCP for inferring sleeping state (e.g. active and quiet sleep) in infants \citep{Howell2019}. Consequently, the influence of different sleep states on the infant FC is still unclear, and capturing respiration during scanning is important for precise sleep state classification and for investigating their potential links with infant neurodevelopment in future research \citep{Mueller2025}.

\section*{Author contributions}

Lingbin Bian (Study conception and design, Material preparation, Data collection, Formal analysis, Methodology,
Software, Validation, Visualization, Writing-original draft, Writing-review $\&$ editing), Feihong Liu (Study conception and design, Material preparation, Data collection,
Writing-review $\&$ editing), Qian Wang (Funding acquisition, Resources,
Writing-review $\&$ editing), Han Zhang (Funding acquisition,
Resources, Writing-review $\&$ editing), and Dinggang Shen
(Study conception and design, Funding acquisition, Methodology, Resources,
Supervision, Writing-review $\&$ editing)

\section*{Funding}

This work was supported in part by the China Ministry of Science and Technology (STI2030-Major Projects-2022ZD0209000, S20240085, STI2030-Major Projects-2022ZD0213100), National Natural Science Foundation of China (grant numbers U23A20295, 82441023, 62131015, 82394432), Shanghai Municipal Central Guided Local Science and Technology Development Fund (No. YDZX20233100001001), The Key R$\&$D Program of Guangdong Province, China (grant number 2023B0303040001), and the Shanghai Pilot Program for Basic Research - Chinese Academy of Science, Shanghai Branch (No. JCYJ-SHFY-2022-014). The computation in this work is also supported by the HPC Platform of ShanghaiTech University.

\section*{Data availability}
Data supporting the current studies findings are available from the corresponding author upon reasonable  request.
\section*{Code availability}
The code of this work is available at: \href{https://github.com/LingbinBian/InfantSCFCModule}{https://github.com/LingbinBian/InfantSCFCModule}. 

\section*{Statements and declarations}
\subsection*{Competing interest}
No competing interest is declared.

\subsection*{Ethical approval}
Ethics approval disclosure statement is not required in this paper. All procedures were approved by the Institutional Review Boards of the University of North Carolina at Chapel Hill and the University of Minnesota.

\subsection*{Consent to participant}

The data used in this paper are from the open source Baby Connectome Project (BCP) \citep*{Howell2019}. Parents of all participants gave the permission before participation. Before the data were collected, families were required to complete all enrollment paperwork including informed consent and HIPAA disclosure, upon arriving for their scheduled visit and MRI scan.

\newpage

%

\newpage

\section*{\Large \textbf{Supplementary information (SI): Robust probabilistic measurement of structural-functional module consistency in infant brain development}}

\section*{SI.1: Correlation analysis of averaged SFMC against age. All subjects (both female and male, gender=0) AP}
\textbf{ROI=100, AP}\\
LH Vis: r = -0.8254, 95\% CI = [0.9505, 0.9904], p-value = 0.0061\\
LH SomMot: r = -0.8377, 95\% CI = [0.8546, 0.9397], p-value = 0.0048\\
LH DorsAttn: r = -0.9491, 95\% CI = [0.8203, 0.8823], p-value = 0.0001\\
LH SalVentAttn: r = -0.9431, 95\% CI = [0.7645, 0.8629], p-value = 0.0001\\
LH Limbic: r = -0.2347, 95\% CI = [0.5928, 0.9623], p-value = 0.5432\\
LH Cont: r = -0.9516, 95\% CI = [0.8285, 0.9126], p-value = 0.0001\\
LH Default: r = -0.9145, 95\% CI = [0.9120, 1.0165], p-value = 0.0006\\
RH Vis: r = -0.5296, 95\% CI = [0.8877, 0.9815], p-value = 0.1425\\
RH SomMot: r = -0.2507, 95\% CI = [0.8068, 0.8904], p-value = 0.5153\\
RH DorsAttn: r = -0.8852, 95\% CI = [0.7341, 0.8308], p-value = 0.0015\\
RH SalVentAttn: r = -0.8182, 95\% CI = [0.7111, 0.8034], p-value = 0.0070\\
RH Limbic: r = -0.2058, 95\% CI = [0.6836, 0.8543], p-value = 0.5952\\
RH Cont: r = -0.7962, 95\% CI = [0.7893, 0.9611], p-value = 0.0102\\
RH Default: r = -0.8903, 95\% CI = [0.8517, 0.9515], p-value = 0.0013\\
\\
\textbf{ROI=200, AP}\\
LH Vis: r = -0.8068, 95\% CI = [0.8253, 0.9022], p-value = 0.0086\\
LH SomMot: r = -0.6575, 95\% CI = [0.7849, 0.9080], p-value = 0.0543\\
LH DorsAttn: r = -0.9469, 95\% CI = [0.7548, 0.8573], p-value = 0.0001\\
LH SalVentAttn: r = -0.9773, 95\% CI = [0.7622, 0.8123], p-value = 0.0000\\
LH Limbic: r = -0.2974, 95\% CI = [0.5820, 0.7314], p-value = 0.4370\\
LH Cont: r = -0.9461, 95\% CI = [0.7708, 0.8937], p-value = 0.0001\\
LH Default: r = -0.9538, 95\% CI = [0.7785, 0.8851], p-value = 0.0001\\
RH Vis: r = -0.7956, 95\% CI = [0.9414, 1.0005], p-value = 0.0103\\
RH SomMot: r = -0.4496, 95\% CI = [0.6357, 0.8396], p-value = 0.2247\\
RH DorsAttn: r = -0.9154, 95\% CI = [0.6381, 0.7466], p-value = 0.0005\\
RH SalVentAttn: r = -0.6788, 95\% CI = [0.5842, 0.7679], p-value = 0.0444\\
RH Limbic: r = -0.6561, 95\% CI = [0.6494, 0.7573], p-value = 0.0550\\
RH Cont: r = -0.8477, 95\% CI = [0.6507, 0.8241], p-value = 0.0039\\
RH Default: r = -0.9379, 95\% CI = [0.7084, 0.7999], p-value = 0.0002\\
\\
\textbf{ROI=300, AP}\\
LH Vis: r = -0.4652, 95\% CI = [0.8706, 0.9630], p-value = 0.2070\\
LH SomMot: r = -0.7388, 95\% CI = [0.6363, 0.7332], p-value = 0.0230\\
LH DorsAttn: r = -0.8902, 95\% CI = [0.6493, 0.7925], p-value = 0.0013\\
LH SalVentAttn: r = -0.9000, 95\% CI = [0.6352, 0.7601], p-value = 0.0009\\
LH Limbic: r = -0.2442, 95\% CI = [0.5488, 0.6848], p-value = 0.5265\\
LH Cont: r = -0.9512, 95\% CI = [0.6865, 0.8011], p-value = 0.0001\\
LH Default: r = -0.9343, 95\% CI = [0.6987, 0.8295], p-value = 0.0002\\
RH Vis: r = -0.5038, 95\% CI = [0.8925, 0.9989], p-value = 0.1667\\
RH SomMot: r = -0.5092, 95\% CI = [0.6461, 0.8049], p-value = 0.1615\\
RH DorsAttn: r = -0.9554, 95\% CI = [0.6542, 0.7399], p-value = 0.0001\\
RH SalVentAttn: r = -0.7630, 95\% CI = [0.5491, 0.6789], p-value = 0.0168\\
RH Limbic: r = -0.8205, 95\% CI = [0.6179, 0.7437], p-value = 0.0067\\
RH Cont: r = -0.8941, 95\% CI = [0.6129, 0.7349], p-value = 0.0011\\
RH Default: r = -0.9704, 95\% CI = [0.6843, 0.7565], p-value = 0.0000\\
\\
\textbf{ROI=400, AP}\\
LH Vis: r = -0.7213, 95\% CI = [0.9105, 0.9612], p-value = 0.0283\\
LH SomMot: r = -0.5858, 95\% CI = [0.5566, 0.6921], p-value = 0.0974\\
LH DorsAttn: r = -0.9334, 95\% CI = [0.6217, 0.7416], p-value = 0.0002\\
LH SalVentAttn: r = -0.8180, 95\% CI = [0.5157, 0.6474], p-value = 0.0071\\
LH Limbic: r = -0.7672, 95\% CI = [0.6075, 0.7370], p-value = 0.0158\\
LH Cont: r = -0.9527, 95\% CI = [0.6054, 0.7047], p-value = 0.0001\\
LH Default: r = -0.9532, 95\% CI = [0.6412, 0.7334], p-value = 0.0001\\
RH Vis: r = -0.8361, 95\% CI = [0.9131, 0.9819], p-value = 0.0050\\
RH SomMot: r = -0.5858, 95\% CI = [0.5633, 0.7081], p-value = 0.0974\\
RH DorsAttn: r = -0.9331, 95\% CI = [0.6014, 0.7162], p-value = 0.0002\\
RH SalVentAttn: r = -0.7058, 95\% CI = [0.5318, 0.6759], p-value = 0.0336\\
RH Limbic: r = -0.6257, 95\% CI = [0.4747, 0.5925], p-value = 0.0715\\
RH Cont: r = -0.8831, 95\% CI = [0.5160, 0.6632], p-value = 0.0016\\
RH Default: r = -0.9838, 95\% CI = [0.5951, 0.6391], p-value = 0.0000\\
\\
\\
\\
\section*{SI.2: All subjects (both female and male, gender=0) PA}

\captionsetup[figure]{labelfont={bf},name={Fig},labelsep=period}
\renewcommand{\thefigure}{SI.2}
\begin{figure*}[!ht]
\centering
\includegraphics[width=0.9\linewidth]{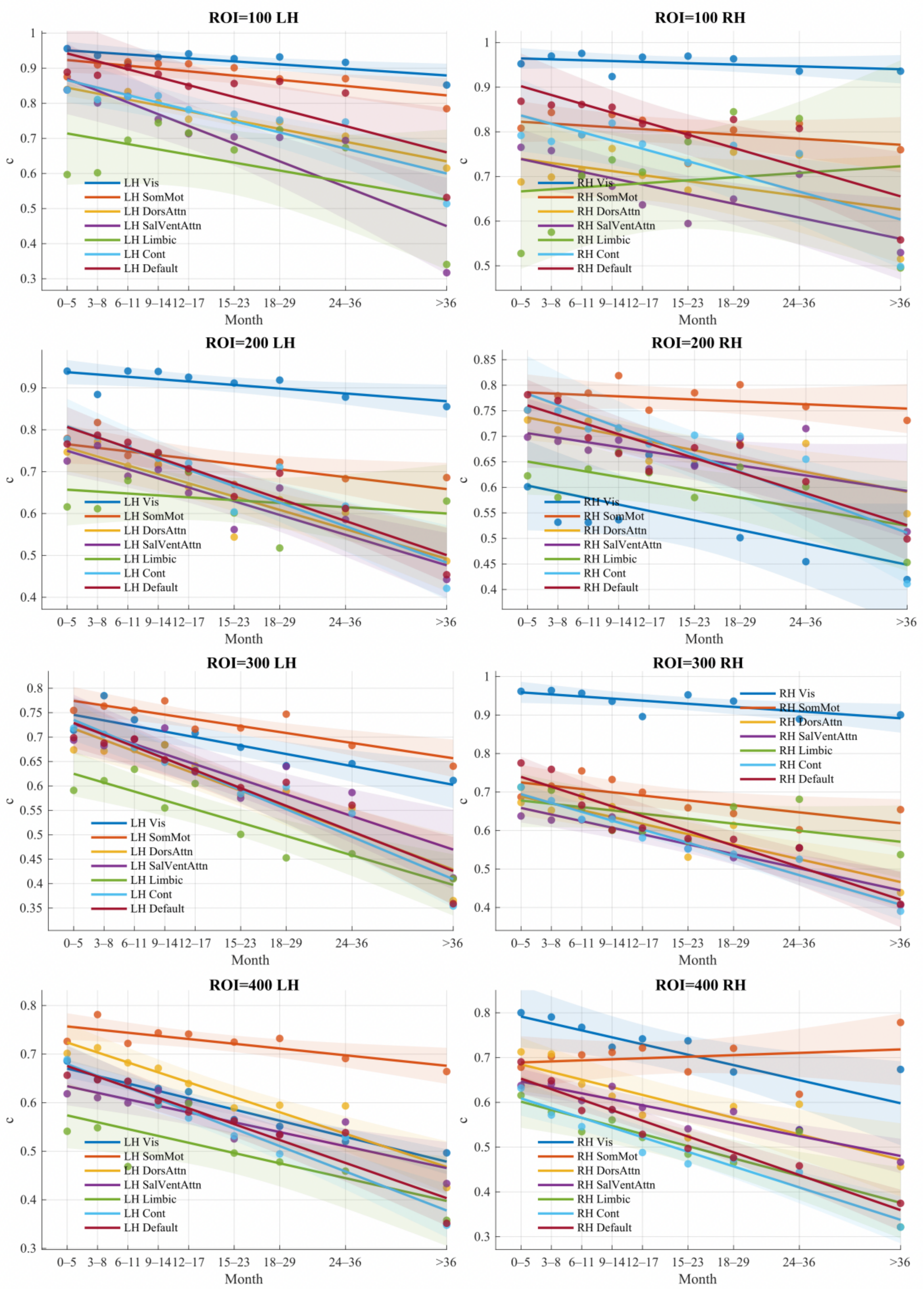}
\caption{\footnotesize \textbf{}}
\label{SI_2}
\end{figure*}

\textbf{ }\\
\textbf{ROI=100, PA}\\
LH Vis: r = -0.7922, 95\% CI = [0.9263, 0.9754], p-value = 0.0109\\
LH SomMot: r = -0.7825, 95\% CI = [0.8876, 0.9591], p-value = 0.0127\\
LH DorsAttn: r = -0.9687, 95\% CI = [0.8206, 0.8686], p-value = 0.0000\\
LH SalVentAttn: r = -0.8807, 95\% CI = [0.7691, 0.9702], p-value = 0.0017\\
LH Limbic: r = -0.4989, 95\% CI = [0.5680, 0.8601], p-value = 0.1716\\
LH Cont: r = -0.8822, 95\% CI = [0.8026, 0.9291], p-value = 0.0016\\
LH Default: r = -0.8020, 95\% CI = [0.8485, 1.0356], p-value = 0.0093\\
RH Vis: r = -0.4220, 95\% CI = [0.9412, 0.9875], p-value = 0.2578\\
RH SomMot: r = -0.4516, 95\% CI = [0.7772, 0.8680], p-value = 0.2223\\
RH DorsAttn: r = -0.4953, 95\% CI = [0.6509, 0.8277], p-value = 0.1752\\
RH SalVentAttn: r = -0.7672, 95\% CI = [0.6725, 0.8058], p-value = 0.0158\\
RH Limbic: r = 0.1430, 95\% CI = [0.4933, 0.8402], p-value = 0.7136\\
RH Cont: r = -0.7903, 95\% CI = [0.7562, 0.9170], p-value = 0.0112\\
RH Default: r = -0.8331, 95\% CI = [0.8293, 0.9753], p-value = 0.0053\\
\\
\textbf{ROI=200, PA}\\
LH Vis: r = -0.7305, 95\% CI = [0.9087, 0.9665], p-value = 0.0254\\
LH SomMot: r = -0.7206, 95\% CI = [0.7196, 0.8114], p-value = 0.0285\\
LH DorsAttn: r = -0.9026, 95\% CI = [0.7009, 0.8138], p-value = 0.0009\\
LH SalVentAttn: r = -0.9062, 95\% CI = [0.6929, 0.8067], p-value = 0.0008\\
LH Limbic: r = -0.2837, 95\% CI = [0.5713, 0.7426], p-value = 0.4594\\
LH Cont: r = -0.9117, 95\% CI = [0.7431, 0.8738], p-value = 0.0006\\
LH Default: r = -0.9437, 95\% CI = [0.7588, 0.8543], p-value = 0.0001\\
RH Vis: r = -0.6192, 95\% CI = [0.5159, 0.6911], p-value = 0.0754\\
RH SomMot: r = -0.3721, 95\% CI = [0.7507, 0.8209], p-value = 0.3240\\
RH DorsAttn: r = -0.8272, 95\% CI = [0.6926, 0.7807], p-value = 0.0059\\
RH SalVentAttn: r = -0.5975, 95\% CI = [0.6385, 0.7737], p-value = 0.0893\\
RH Limbic: r = -0.6545, 95\% CI = [0.5858, 0.7152], p-value = 0.0558\\
RH Cont: r = -0.8544, 95\% CI = [0.7093, 0.8567], p-value = 0.0034\\
RH Default: r = -0.9015, 95\% CI = [0.7103, 0.8107], p-value = 0.0009\\
\\
\textbf{ROI=300, PA}\\
LH Vis: r = -0.8771, 95\% CI = [0.7107, 0.7805], p-value = 0.0019\\
LH SomMot: r = -0.8807, 95\% CI = [0.7460, 0.8023], p-value = 0.0017\\
LH DorsAttn: r = -0.9268, 95\% CI = [0.6651, 0.7694], p-value = 0.0003\\
LH SalVentAttn: r = -0.8864, 95\% CI = [0.6669, 0.7866], p-value = 0.0015\\
LH Limbic: r = -0.9111, 95\% CI = [0.5791, 0.6708], p-value = 0.0006\\
LH Cont: r = -0.9594, 95\% CI = [0.6924, 0.7778], p-value = 0.0000\\
LH Default: r = -0.9354, 95\% CI = [0.6780, 0.7802], p-value = 0.0002\\
RH Vis: r = -0.7393, 95\% CI = [0.9316, 0.9863], p-value = 0.0228\\
RH SomMot: r = -0.7186, 95\% CI = [0.6792, 0.7703], p-value = 0.0292\\
RH DorsAttn: r = -0.8933, 95\% CI = [0.6387, 0.7384], p-value = 0.0012\\
RH SalVentAttn: r = -0.9301, 95\% CI = [0.6210, 0.6965], p-value = 0.0003\\
RH Limbic: r = -0.5684, 95\% CI = [0.6089, 0.7471], p-value = 0.1103\\
RH Cont: r = -0.9756, 95\% CI = [0.6657, 0.7232], p-value = 0.0000\\
RH Default: r = -0.9380, 95\% CI = [0.6869, 0.7919], p-value = 0.0002\\
\\
\textbf{ROI=400, PA}\\
LH Vis: r = -0.9770, 95\% CI = [0.6517, 0.6890], p-value = 0.0000\\
LH SomMot: r = -0.7994, 95\% CI = [0.7298, 0.7841], p-value = 0.0097\\
LH DorsAttn: r = -0.9468, 95\% CI = [0.6851, 0.7627], p-value = 0.0001\\
LH SalVentAttn: r = -0.8866, 95\% CI = [0.5947, 0.6730], p-value = 0.0014\\
LH Limbic: r = -0.7582, 95\% CI = [0.5063, 0.6414], p-value = 0.0179\\
LH Cont: r = -0.9579, 95\% CI = [0.6391, 0.7196], p-value = 0.0000\\
LH Default: r = -0.9461, 95\% CI = [0.6340, 0.7170], p-value = 0.0001\\
RH Vis: r = -0.7827, 95\% CI = [0.7231, 0.8603], p-value = 0.0126\\
RH SomMot: r = 0.2163, 95\% CI = [0.6306, 0.7477], p-value = 0.5762\\
RH DorsAttn: r = -0.8467, 95\% CI = [0.6252, 0.7446], p-value = 0.0040\\
RH SalVentAttn: r = -0.9354, 95\% CI = [0.6187, 0.6748], p-value = 0.0002\\
RH Limbic: r = -0.8659, 95\% CI = [0.5434, 0.6597], p-value = 0.0025\\
RH Cont: r = -0.9534, 95\% CI = [0.5707, 0.6471], p-value = 0.0001\\
RH Default: r = -0.9688, 95\% CI = [0.6195, 0.6864], p-value = 0.0000\\

\section*{SI.3: Female (gender=1) AP}
\captionsetup[figure]{labelfont={bf},name={Fig },labelsep=period}
\renewcommand{\thefigure}{SI.3}
\begin{figure*}[!ht]
\centering
\includegraphics[width=0.9\linewidth]{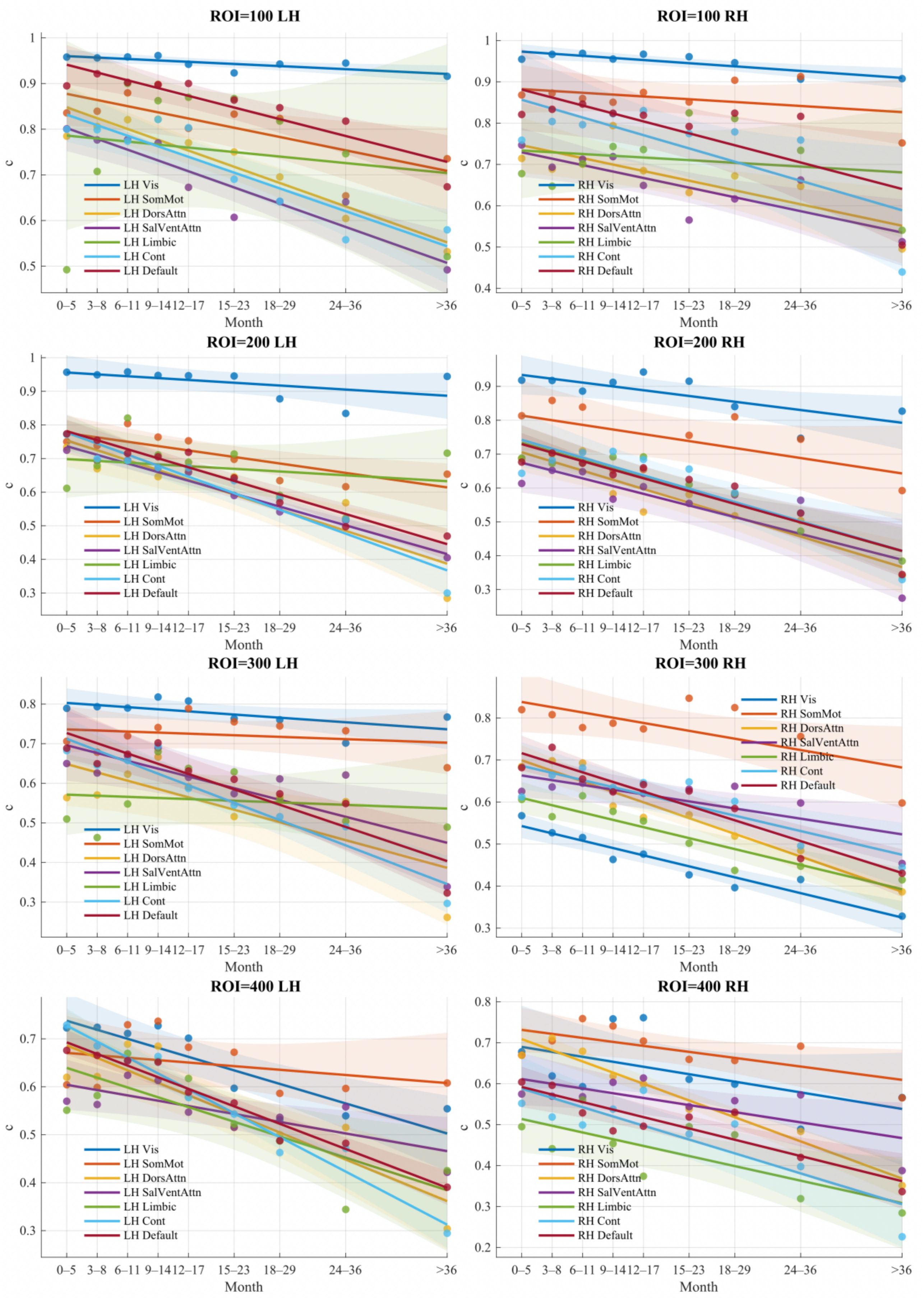}
\caption{\footnotesize \textbf{}}
\label{SI_3}
\end{figure*}
\textbf{ }\\
\textbf{ROI=100, AP}\\
LH Vis: r = -0.7902, 95\% CI = [0.9465, 0.9735], p-value = 0.0113\\
LH SomMot: r = -0.7350, 95\% CI = [0.8081, 0.9467], p-value = 0.0241\\
LH DorsAttn: r = -0.9421, 95\% CI = [0.8009, 0.8946], p-value = 0.0001\\
LH SalVentAttn: r = -0.9345, 95\% CI = [0.7523, 0.8525], p-value = 0.0002\\
LH Limbic: r = -0.1744, 95\% CI = [0.5794, 0.9927], p-value = 0.6536\\
LH Cont: r = -0.9082, 95\% CI = [0.7726, 0.8909], p-value = 0.0007\\
LH Default: r = -0.9111, 95\% CI = [0.8982, 0.9840], p-value = 0.0006\\
RH Vis: r = -0.8507, 95\% CI = [0.9557, 0.9908], p-value = 0.0036\\
RH SomMot: r = -0.3951, 95\% CI = [0.8247, 0.9403], p-value = 0.2926\\
RH DorsAttn: r = -0.7938, 95\% CI = [0.6799, 0.8131], p-value = 0.0106\\
RH SalVentAttn: r = -0.8264, 95\% CI = [0.6704, 0.7885], p-value = 0.0060\\
RH Limbic: r = -0.2006, 95\% CI = [0.6183, 0.8491], p-value = 0.6048\\
RH Cont: r = -0.7284, 95\% CI = [0.7444, 0.9688], p-value = 0.0261\\
RH Default: r = -0.7350, 95\% CI = [0.7823, 0.9803], p-value = 0.0241\\
\\
\textbf{ROI=200, AP}\\
LH Vis: r = -0.5247, 95\% CI = [0.9059, 1.0060], p-value = 0.1469\\
LH SomMot: r = -0.8106, 95\% CI = [0.7238, 0.8279], p-value = 0.0080\\
LH DorsAttn: r = -0.9027, 95\% CI = [0.6759, 0.8317], p-value = 0.0009\\
LH SalVentAttn: r = -0.9893, 95\% CI = [0.7156, 0.7579], p-value = 0.0000\\
LH Limbic: r = -0.2474, 95\% CI = [0.5835, 0.8134], p-value = 0.5210\\
LH Cont: r = -0.9607, 95\% CI = [0.7244, 0.8300], p-value = 0.0000\\
LH Default: r = -0.9754, 95\% CI = [0.7475, 0.8152], p-value = 0.0000\\
RH Vis: r = -0.7381, 95\% CI = [0.8763, 0.9914], p-value = 0.0232\\
RH SomMot: r = -0.5926, 95\% CI = [0.7102, 0.9171], p-value = 0.0927\\
RH DorsAttn: r = -0.9353, 95\% CI = [0.6483, 0.7627], p-value = 0.0002\\
RH SalVentAttn: r = -0.8236, 95\% CI = [0.5864, 0.7623], p-value = 0.0064\\
RH Limbic: r = -0.9424, 95\% CI = [0.6828, 0.7841], p-value = 0.0001\\
RH Cont: r = -0.8775, 95\% CI = [0.6622, 0.8219], p-value = 0.0019\\
RH Default: r = -0.9284, 95\% CI = [0.6733, 0.7858], p-value = 0.0003\\
\\
\textbf{ROI=300, AP}\\
LH Vis: r = -0.6303, 95\% CI = [0.7663, 0.8393], p-value = 0.0688\\
LH SomMot: r = -0.2483, 95\% CI = [0.6777, 0.7943], p-value = 0.5194\\
LH DorsAttn: r = -0.7430, 95\% CI = [0.5433, 0.7535], p-value = 0.0218\\
LH SalVentAttn: r = -0.7765, 95\% CI = [0.6070, 0.7851], p-value = 0.0139\\
LH Limbic: r = -0.1528, 95\% CI = [0.4698, 0.6730], p-value = 0.6946\\
LH Cont: r = -0.9515, 95\% CI = [0.6594, 0.7651], p-value = 0.0001\\
LH Default: r = -0.9104, 95\% CI = [0.6614, 0.7923], p-value = 0.0006\\
RH Vis: r = -0.9591, 95\% CI = [0.5141, 0.5714], p-value = 0.0000\\
RH SomMot: r = -0.6968, 95\% CI = [0.7665, 0.9097], p-value = 0.0370\\
RH DorsAttn: r = -0.9689, 95\% CI = [0.6636, 0.7347], p-value = 0.0000\\
RH SalVentAttn: r = -0.7689, 95\% CI = [0.6112, 0.7151], p-value = 0.0154\\
RH Limbic: r = -0.9307, 95\% CI = [0.5716, 0.6477], p-value = 0.0003\\
RH Cont: r = -0.8600, 95\% CI = [0.6301, 0.7421], p-value = 0.0029\\
RH Default: r = -0.9504, 95\% CI = [0.6746, 0.7578], p-value = 0.0001\\
\\
\textbf{ROI=400, AP}
LH Vis: r = -0.8752, 95\% CI = [0.6798, 0.7956], p-value = 0.0020\\
LH SomMot: r = -0.3432, 95\% CI = [0.5939, 0.7476], p-value = 0.3659\\
LH DorsAttn: r = -0.8907, 95\% CI = [0.6125, 0.7602], p-value = 0.0013\\
LH SalVentAttn: r = -0.7630, 95\% CI = [0.5520, 0.6565], p-value = 0.0168\\
LH Limbic: r = -0.7739, 95\% CI = [0.5463, 0.7325], p-value = 0.0144\\
LH Cont: r = -0.9821, 95\% CI = [0.6920, 0.7630], p-value = 0.0000\\
LH Default: r = -0.9821, 95\% CI = [0.6668, 0.7186], p-value = 0.0000\\
RH Vis: r = -0.5573, 95\% CI = [0.5893, 0.7904], p-value = 0.1190\\
RH SomMot: r = -0.7063, 95\% CI = [0.6769, 0.7859], p-value = 0.0334\\
RH DorsAttn: r = -0.9778, 95\% CI = [0.6763, 0.7413], p-value = 0.0000\\
RH SalVentAttn: r = -0.7102, 95\% CI = [0.5476, 0.6750], p-value = 0.0320\\
RH Limbic: r = -0.7439, 95\% CI = [0.4316, 0.5960], p-value = 0.0215\\
RH Cont: r = -0.8546, 95\% CI = [0.5115, 0.6643], p-value = 0.0033\\
RH Default: r = -0.9006, 95\% CI = [0.5425, 0.6415], p-value = 0.0009\\

\section*{SI.4: Female (gender=1) PA}
\captionsetup[figure]{labelfont={bf},name={Fig },labelsep=period}
\renewcommand{\thefigure}{SI.4}
\begin{figure*}[!ht]
\centering
\includegraphics[width=0.9\linewidth]{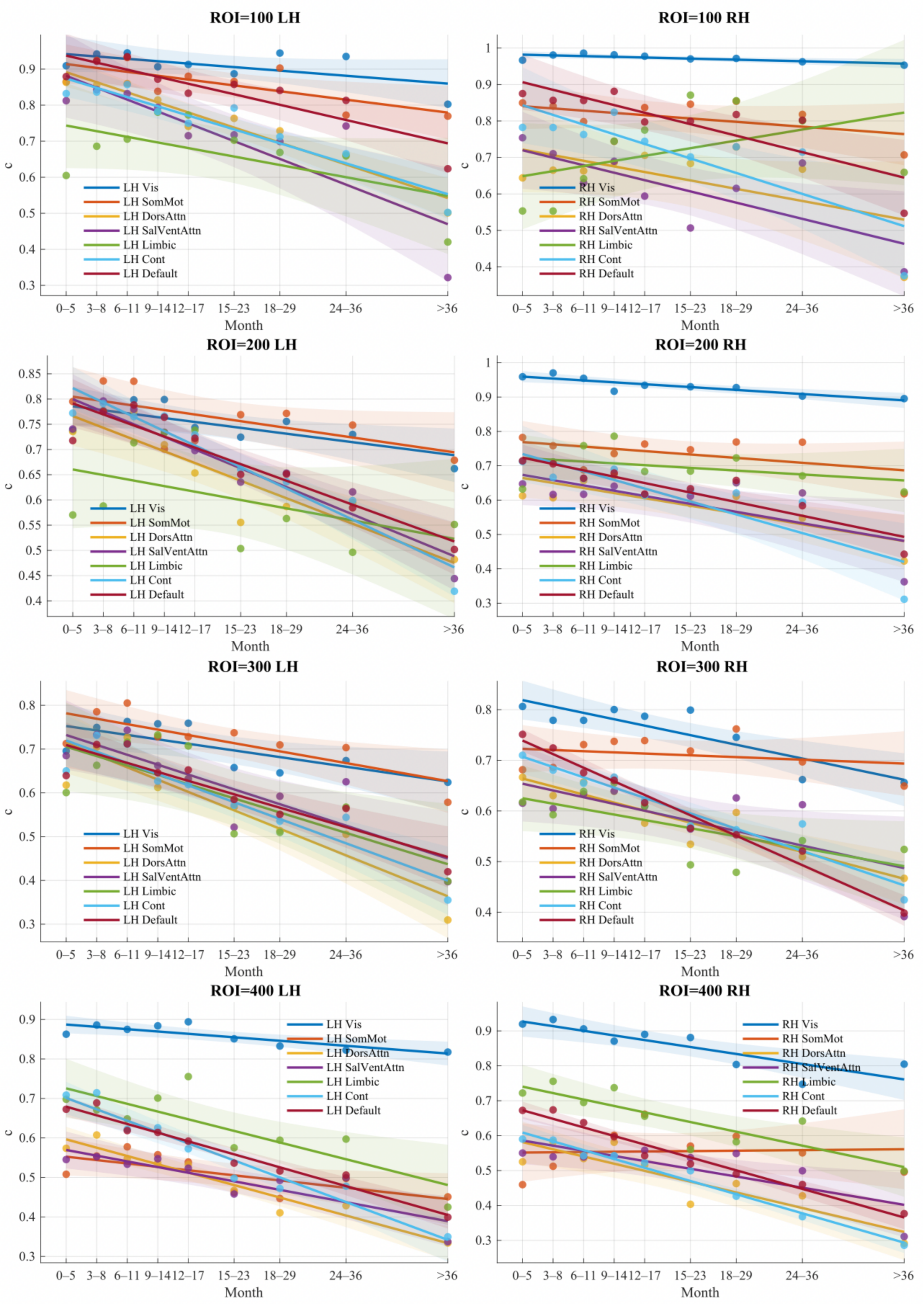}
\caption{\footnotesize \textbf{}}
\label{SI_4}
\end{figure*}
\textbf{ }\\
\textbf{ROI=100, PA}\\
LH Vis: r = -0.5938, 95\% CI = [0.8922, 0.9910], p-value = 0.0919\\
LH SomMot: r = -0.7419, 95\% CI = [0.8598, 0.9682], p-value = 0.0221\\
LH DorsAttn: r = -0.9675, 95\% CI = [0.8509, 0.9322], p-value = 0.0000\\
LH SalVentAttn: r = -0.8455, 95\% CI = [0.7655, 0.9971], p-value = 0.0041\\
LH Limbic: r = -0.5941, 95\% CI = [0.6255, 0.8614], p-value = 0.0916\\
LH Cont: r = -0.9430, 95\% CI = [0.8234, 0.9243], p-value = 0.0001\\
LH Default: r = -0.8735, 95\% CI = [0.8769, 0.9979], p-value = 0.0021\\
RH Vis: r = -0.7596, 95\% CI = [0.9725, 0.9913], p-value = 0.0176\\
RH SomMot: r = -0.4794, 95\% CI = [0.7782, 0.9030], p-value = 0.1916\\
RH DorsAttn: r = -0.5908, 95\% CI = [0.6046, 0.8386], p-value = 0.0939\\
RH SalVentAttn: r = -0.7356, 95\% CI = [0.6145, 0.8251], p-value = 0.0239\\
RH Limbic: r = 0.4711, 95\% CI = [0.5029, 0.7942], p-value = 0.2005\\
RH Cont: r = -0.8180, 95\% CI = [0.7393, 0.9471], p-value = 0.0070\\
RH Default: r = -0.8384, 95\% CI = [0.8306, 0.9823], p-value = 0.0047\\
\\
\textbf{ROI=200, PA}\\
LH Vis: r = -0.7448, 95\% CI = [0.7469, 0.8244], p-value = 0.0213\\
LH SomMot: r = -0.6451, 95\% CI = [0.7464, 0.8629], p-value = 0.0606\\
LH DorsAttn: r = -0.9122, 95\% CI = [0.7080, 0.8246], p-value = 0.0006\\
LH SalVentAttn: r = -0.9423, 95\% CI = [0.7504, 0.8491], p-value = 0.0001\\
LH Limbic: r = -0.4659, 95\% CI = [0.5440, 0.7769], p-value = 0.2062\\
LH Cont: r = -0.9662, 95\% CI = [0.7794, 0.8639], p-value = 0.0000\\
LH Default: r = -0.9309, 95\% CI = [0.7435, 0.8393], p-value = 0.0003\\
RH Vis: r = -0.8869, 95\% CI = [0.9433, 0.9754], p-value = 0.0014\\
RH SomMot: r = -0.5152, 95\% CI = [0.7079, 0.8302], p-value = 0.1558\\
RH DorsAttn: r = -0.8209, 95\% CI = [0.6076, 0.7216], p-value = 0.0067\\
RH SalVentAttn: r = -0.6986, 95\% CI = [0.5859, 0.7612], p-value = 0.0363\\
RH Limbic: r = -0.3966, 95\% CI = [0.6551, 0.7903], p-value = 0.2906\\
RH Cont: r = -0.8621, 95\% CI = [0.6519, 0.8163], p-value = 0.0028\\
RH Default: r = -0.9111, 95\% CI = [0.6770, 0.7700], p-value = 0.0006\\
\\
\textbf{ROI=300, PA}\\
LH Vis: r = -0.7479, 95\% CI = [0.7026, 0.8027], p-value = 0.0205\\
LH SomMot: r = -0.7937, 95\% CI = [0.7288, 0.8346], p-value = 0.0107\\
LH DorsAttn: r = -0.9128, 95\% CI = [0.6445, 0.7844], p-value = 0.0006\\
LH SalVentAttn: r = -0.8470, 95\% CI = [0.6532, 0.8111], p-value = 0.0040\\
LH Limbic: r = -0.7632, 95\% CI = [0.6042, 0.8067], p-value = 0.0167\\
LH Cont: r = -0.9287, 95\% CI = [0.6626, 0.7766], p-value = 0.0003\\
LH Default: r = -0.9123, 95\% CI = [0.6582, 0.7605], p-value = 0.0006\\
RH Vis: r = -0.8779, 95\% CI = [0.7807, 0.8572], p-value = 0.0019\\
RH SomMot: r = -0.2676, 95\% CI = [0.6760, 0.7695], p-value = 0.4863\\
RH DorsAttn: r = -0.9221, 95\% CI = [0.6271, 0.7011], p-value = 0.0004\\
RH SalVentAttn: r = -0.7063, 95\% CI = [0.5794, 0.7283], p-value = 0.0334\\
RH Limbic: r = -0.6776, 95\% CI = [0.5602, 0.6900], p-value = 0.0449\\
RH Cont: r = -0.9578, 95\% CI = [0.6731, 0.7411], p-value = 0.0000\\
RH Default: r = -0.9893, 95\% CI = [0.7171, 0.7612], p-value = 0.0000\\
\\
\textbf{ROI=400, PA}\\
LH Vis: r = -0.8295, 95\% CI = [0.8651, 0.9092], p-value = 0.0057\\
LH SomMot: r = -0.7070, 95\% CI = [0.5047, 0.6003], p-value = 0.0332\\
LH DorsAttn: r = -0.9651, 95\% CI = [0.5646, 0.6281], p-value = 0.0000\\
LH SalVentAttn: r = -0.8647, 95\% CI = [0.5229, 0.6162], p-value = 0.0026\\
LH Limbic: r = -0.8251, 95\% CI = [0.6509, 0.8002], p-value = 0.0062\\
LH Cont: r = -0.9693, 95\% CI = [0.6605, 0.7418], p-value = 0.0000\\
LH Default: r = -0.9795, 95\% CI = [0.6541, 0.7043], p-value = 0.0000\\
RH Vis: r = -0.8644, 95\% CI = [0.8839, 0.9699], p-value = 0.0026\\
RH SomMot: r = 0.0505, 95\% CI = [0.4678, 0.6353], p-value = 0.8973\\
RH DorsAttn: r = -0.8900, 95\% CI = [0.5227, 0.6402], p-value = 0.0013\\
RH SalVentAttn: r = -0.7721, 95\% CI = [0.5185, 0.6537], p-value = 0.0148\\
RH Limbic: r = -0.8555, 95\% CI = [0.6782, 0.8025], p-value = 0.0033\\
RH Cont: r = -0.9885, 95\% CI = [0.5874, 0.6304], p-value = 0.0000\\
RH Default: r = -0.9842, 95\% CI = [0.6481, 0.6974], p-value = 0.0000\\

\section*{SI.5: Male (gender=2) AP}
\captionsetup[figure]{labelfont={bf},name={Fig },labelsep=period}
\renewcommand{\thefigure}{SI.5}
\begin{figure*}[!ht]
\centering
\includegraphics[width=0.9\linewidth]{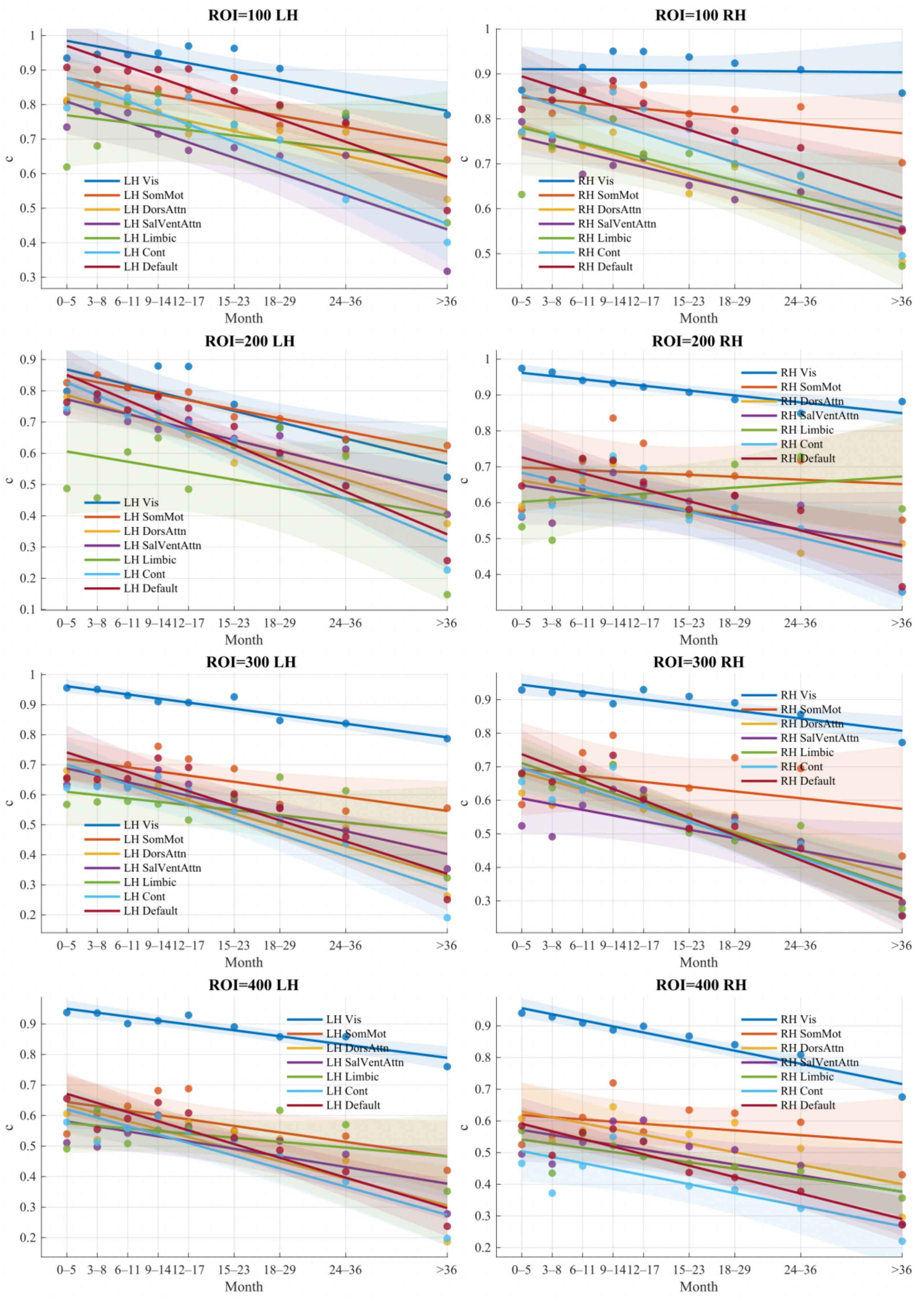}
\caption{\footnotesize \textbf{}}
\label{SI_5}
\end{figure*}
\textbf{ }\\
\textbf{ROI=100, AP}\\
LH Vis: r = -0.8231, 95\% CI = [0.9224, 1.0469], p-value = 0.0064\\
LH SomMot: r = -0.8171, 95\% CI = [0.8154, 0.9371], p-value = 0.0072\\
LH DorsAttn: r = -0.9026, 95\% CI = [0.7786, 0.8822], p-value = 0.0009\\
LH SalVentAttn: r = -0.8678, 95\% CI = [0.7143, 0.9032], p-value = 0.0024\\
LH Limbic: r = -0.3730, 95\% CI = [0.6206, 0.9178], p-value = 0.3229\\
LH Cont: r = -0.9211, 95\% CI = [0.7985, 0.9581], p-value = 0.0004\\
LH Default: r = -0.9119, 95\% CI = [0.8940, 1.0460], p-value = 0.0006\\
RH Vis: r = -0.0622, 95\% CI = [0.8597, 0.9617], p-value = 0.8737\\
RH SomMot: r = -0.4770, 95\% CI = [0.7823, 0.9133], p-value = 0.1942\\
RH DorsAttn: r = -0.9077, 95\% CI = [0.7330, 0.8375], p-value = 0.0007\\
RH SalVentAttn: r = -0.9240, 95\% CI = [0.7199, 0.7950], p-value = 0.0004\\
RH Limbic: r = -0.6608, 95\% CI = [0.6744, 0.8859], p-value = 0.0526\\
RH Cont: r = -0.8130, 95\% CI = [0.7678, 0.9403], p-value = 0.0077\\
RH Default: r = -0.8791, 95\% CI = [0.8292, 0.9600], p-value = 0.0018\\
\\
\textbf{ROI=200, AP}\\
LH Vis: r = -0.8476, 95\% CI = [0.7841, 0.9521], p-value = 0.0039\\
LH SomMot: r = -0.9682, 95\% CI = [0.8190, 0.8745], p-value = 0.0000\\
LH DorsAttn: r = -0.9502, 95\% CI = [0.7327, 0.8405], p-value = 0.0001\\
LH SalVentAttn: r = -0.9110, 95\% CI = [0.7131, 0.8324], p-value = 0.0006\\
LH Limbic: r = -0.4124, 95\% CI = [0.4038, 0.8078], p-value = 0.2700\\
LH Cont: r = -0.9503, 95\% CI = [0.7520, 0.9003], p-value = 0.0001\\
LH Default: r = -0.9435, 95\% CI = [0.7712, 0.9310], p-value = 0.0001\\
RH Vis: r = -0.9025, 95\% CI = [0.9378, 0.9856], p-value = 0.0009\\
RH SomMot: r = -0.1653, 95\% CI = [0.5744, 0.8218], p-value = 0.6709\\
RH DorsAttn: r = -0.7470, 95\% CI = [0.5880, 0.7342], p-value = 0.0207\\
RH SalVentAttn: r = -0.5622, 95\% CI = [0.5389, 0.7576], p-value = 0.1151\\
RH Limbic: r = 0.2537, 95\% CI = [0.4815, 0.7227], p-value = 0.5101\\
RH Cont: r = -0.7107, 95\% CI = [0.5746, 0.7925], p-value = 0.0319\\
RH Default: r = -0.8445, 95\% CI = [0.6477, 0.8048], p-value = 0.0042\\
\\
\textbf{ROI=300, AP}\\
LH Vis: r = -0.9599, 95\% CI = [0.9394, 0.9836], p-value = 0.0000\\
LH SomMot: r = -0.7276, 95\% CI = [0.6467, 0.7920], p-value = 0.0263\\
LH DorsAttn: r = -0.9546, 95\% CI = [0.6481, 0.7509], p-value = 0.0001\\
LH SalVentAttn: r = -0.9128, 95\% CI = [0.6309, 0.7446], p-value = 0.0006\\
LH Limbic: r = -0.4757, 95\% CI = [0.4955, 0.7243], p-value = 0.1956\\
LH Cont: r = -0.9190, 95\% CI = [0.6210, 0.7801], p-value = 0.0005\\
LH Default: r = -0.8958, 95\% CI = [0.6518, 0.8305], p-value = 0.0011\\
RH Vis: r = -0.8855, 95\% CI = [0.9124, 0.9765], p-value = 0.0015\\
RH SomMot: r = -0.3575, 95\% CI = [0.5551, 0.8305], p-value = 0.3449\\
RH DorsAttn: r = -0.8690, 95\% CI = [0.6025, 0.7629], p-value = 0.0024\\
RH SalVentAttn: r = -0.6754, 95\% CI = [0.5025, 0.7086], p-value = 0.0459\\
RH Limbic: r = -0.9117, 95\% CI = [0.6355, 0.7865], p-value = 0.0006\\
RH Cont: r = -0.9110, 95\% CI = [0.6225, 0.7698], p-value = 0.0006\\
RH Default: r = -0.9375, 95\% CI = [0.6662, 0.8088], p-value = 0.0002\\
\\
\textbf{ROI=400, AP}\\
LH Vis: r = -0.9327, 95\% CI = [0.9221, 0.9774], p-value = 0.0002\\
LH SomMot: r = -0.6793, 95\% CI = [0.5586, 0.7303], p-value = 0.0442\\
LH DorsAttn: r = -0.8369, 95\% CI = [0.5387, 0.7300], p-value = 0.0049\\
LH SalVentAttn: r = -0.7270, 95\% CI = [0.4949, 0.6666], p-value = 0.0265\\
LH Limbic: r = -0.4435, 95\% CI = [0.4767, 0.6744], p-value = 0.2319\\
LH Cont: r = -0.9000, 95\% CI = [0.5460, 0.6954], p-value = 0.0009\\
LH Default: r = -0.9248, 95\% CI = [0.6025, 0.7396], p-value = 0.0004\\
RH Vis: r = -0.9607, 95\% CI = [0.9250, 0.9867], p-value = 0.0000\\
RH SomMot: r = -0.3510, 95\% CI = [0.5154, 0.7216], p-value = 0.3544\\
RH DorsAttn: r = -0.7476, 95\% CI = [0.5387, 0.7201], p-value = 0.0206\\
RH SalVentAttn: r = -0.6541, 95\% CI = [0.4709, 0.6721], p-value = 0.0560\\
RH Limbic: r = -0.8028, 95\% CI = [0.4869, 0.5954], p-value = 0.0092\\
RH Cont: r = -0.7417, 95\% CI = [0.4091, 0.6004], p-value = 0.0222\\
RH Default: r = -0.9311, 95\% CI = [0.5385, 0.6435], p-value = 0.0003\\

\section*{SI.6: Male (gender=2) PA}
\captionsetup[figure]{labelfont={bf},name={Fig },labelsep=period}
\renewcommand{\thefigure}{SI.6}
\begin{figure*}[!ht]
\centering
\includegraphics[width=0.9\linewidth]{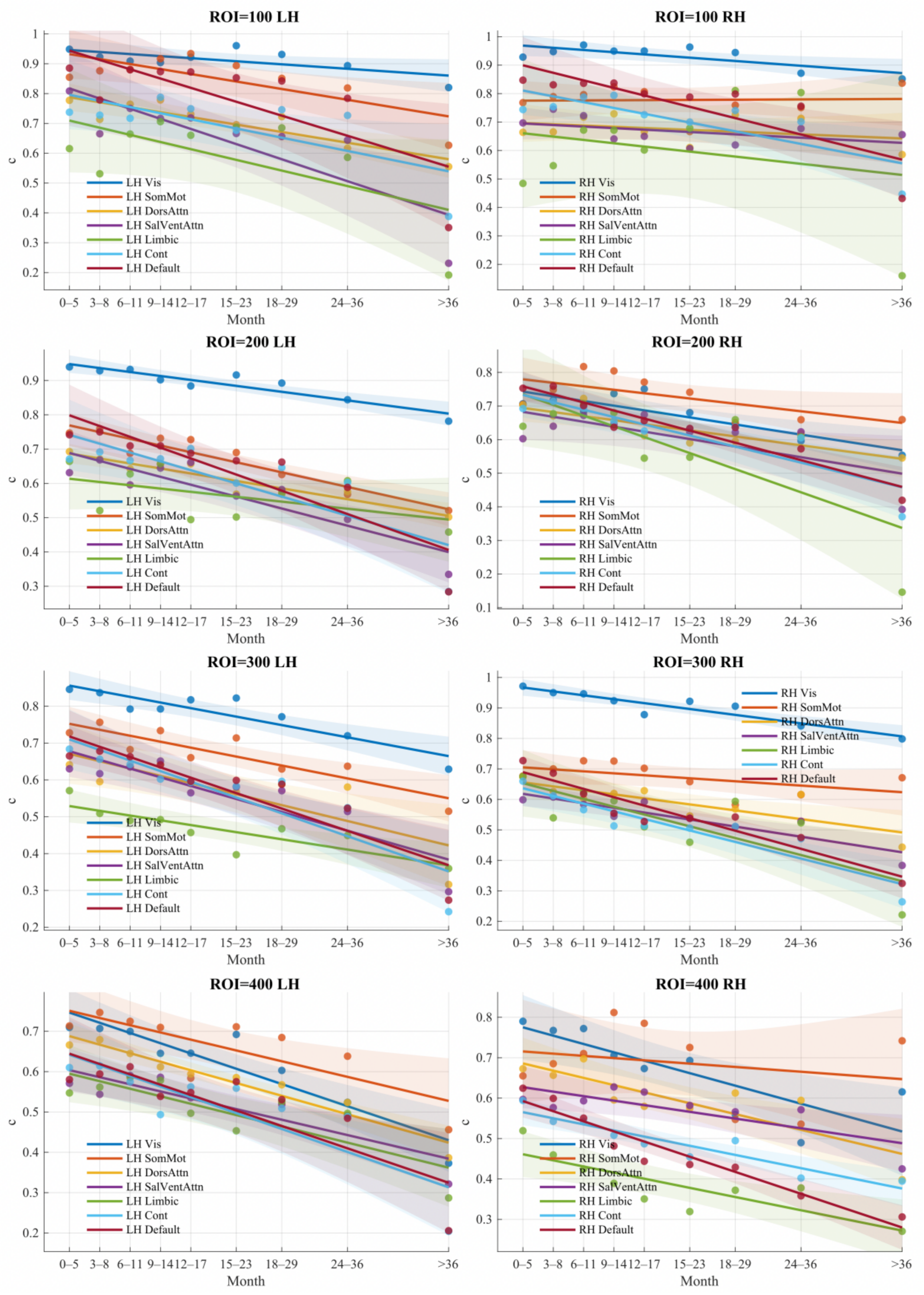}
\caption{\footnotesize \textbf{}}
\label{SI_6}
\end{figure*}
\textbf{ }\\
\textbf{ROI=100, PA}\\
LH Vis: r = -0.6916, 95\% CI = [0.9063, 0.9862], p-value = 0.0390\\
LH SomMot: r = -0.7501, 95\% CI = [0.8504, 1.0146], p-value = 0.0199\\
LH DorsAttn: r = -0.8976, 95\% CI = [0.7423, 0.8332], p-value = 0.0010\\
LH SalVentAttn: r = -0.8324, 95\% CI = [0.6915, 0.9430], p-value = 0.0054\\
LH Limbic: r = -0.6094, 95\% CI = [0.5360, 0.8832], p-value = 0.0815\\
LH Cont: r = -0.7036, 95\% CI = [0.6805, 0.9114], p-value = 0.0344\\
LH Default: r = -0.7466, 95\% CI = [0.7894, 1.0983], p-value = 0.0208\\
RH Vis: r = -0.7638, 95\% CI = [0.9323, 1.0051], p-value = 0.0166\\
RH SomMot: r = 0.0384, 95\% CI = [0.7062, 0.8442], p-value = 0.9218\\
RH DorsAttn: r = -0.3430, 95\% CI = [0.6305, 0.7623], p-value = 0.3662\\
RH SalVentAttn: r = -0.4802, 95\% CI = [0.6396, 0.7506], p-value = 0.1908\\
RH Limbic: r = -0.2419, 95\% CI = [0.3989, 0.9227], p-value = 0.5306\\
RH Cont: r = -0.8000, 95\% CI = [0.7253, 0.8963], p-value = 0.0096\\
RH Default: r = -0.8315, 95\% CI = [0.8009, 0.9984], p-value = 0.0055\\
\\
\textbf{ROI=200, PA}\\
LH Vis: r = -0.9298, 95\% CI = [0.9225, 0.9734], p-value = 0.0003\\
LH SomMot: r = -0.9498, 95\% CI = [0.7333, 0.8051], p-value = 0.0001\\
LH DorsAttn: r = -0.9458, 95\% CI = [0.6593, 0.7149], p-value = 0.0001\\
LH SalVentAttn: r = -0.8933, 95\% CI = [0.6239, 0.7534], p-value = 0.0012\\
LH Limbic: r = -0.5100, 95\% CI = [0.5239, 0.7032], p-value = 0.1607\\
LH Cont: r = -0.8130, 95\% CI = [0.6389, 0.8441], p-value = 0.0077\\
LH Default: r = -0.8911, 95\% CI = [0.7094, 0.8877], p-value = 0.0013\\
RH Vis: r = -0.8666, 95\% CI = [0.6981, 0.7877], p-value = 0.0025\\
RH SomMot: r = -0.6663, 95\% CI = [0.7148, 0.8441], p-value = 0.0500\\
RH DorsAttn: r = -0.8587, 95\% CI = [0.6540, 0.7345], p-value = 0.0030\\
RH SalVentAttn: r = -0.7032, 95\% CI = [0.5998, 0.7655], p-value = 0.0346\\
RH Limbic: r = -0.7540, 95\% CI = [0.5806, 0.8895], p-value = 0.0189\\
RH Cont: r = -0.8773, 95\% CI = [0.6661, 0.8000], p-value = 0.0019\\
RH Default: r = -0.9508, 95\% CI = [0.7150, 0.8019], p-value = 0.0001\\
\\
\textbf{ROI=300, PA}\\
LH Vis: r = -0.9111, 95\% CI = [0.8175, 0.8947], p-value = 0.0006\\
LH SomMot: r = -0.8907, 95\% CI = [0.7064, 0.7983], p-value = 0.0013\\
LH DorsAttn: r = -0.8070, 95\% CI = [0.5897, 0.7518], p-value = 0.0086\\
LH SalVentAttn: r = -0.8837, 95\% CI = [0.6084, 0.7473], p-value = 0.0016\\
LH Limbic: r = -0.8495, 95\% CI = [0.4845, 0.5737], p-value = 0.0037\\
LH Cont: r = -0.8909, 95\% CI = [0.6286, 0.7913], p-value = 0.0013\\
LH Default: r = -0.9105, 95\% CI = [0.6473, 0.7887], p-value = 0.0006\\
RH Vis: r = -0.9360, 95\% CI = [0.9401, 0.9939], p-value = 0.0002\\
RH SomMot: r = -0.5446, 95\% CI = [0.6491, 0.7611], p-value = 0.1295\\
RH DorsAttn: r = -0.8063, 95\% CI = [0.6021, 0.7096], p-value = 0.0087\\
RH SalVentAttn: r = -0.9053, 95\% CI = [0.5781, 0.6583], p-value = 0.0008\\
RH Limbic: r = -0.7941, 95\% CI = [0.5435, 0.7627], p-value = 0.0106\\
RH Cont: r = -0.9256, 95\% CI = [0.5791, 0.6937], p-value = 0.0003\\
RH Default: r = -0.9464, 95\% CI = [0.6372, 0.7415], p-value = 0.0001\\
\\
\textbf{ROI=400, PA}\\
LH Vis: r = -0.9271, 95\% CI = [0.6894, 0.8034], p-value = 0.0003\\
LH SomMot: r = -0.7898, 95\% CI = [0.6735, 0.8281], p-value = 0.0113\\
LH DorsAttn: r = -0.9673, 95\% CI = [0.6571, 0.7189], p-value = 0.0000\\
LH SalVentAttn: r = -0.8761, 95\% CI = [0.5499, 0.6576], p-value = 0.0019\\
LH Limbic: r = -0.8268, 95\% CI = [0.5243, 0.6656], p-value = 0.0060\\
LH Cont: r = -0.8588, 95\% CI = [0.5545, 0.7292], p-value = 0.0030\\
LH Default: r = -0.8444, 95\% CI = [0.5540, 0.7349], p-value = 0.0042\\
RH Vis: r = -0.8212, 95\% CI = [0.6953, 0.8549], p-value = 0.0067\\
RH SomMot: r = -0.2328, 95\% CI = [0.5880, 0.8430], p-value = 0.5466\\
RH DorsAttn: r = -0.8430, 95\% CI = [0.6222, 0.7495], p-value = 0.0043\\
RH SalVentAttn: r = -0.7666, 95\% CI = [0.5753, 0.6788], p-value = 0.0159\\
RH Limbic: r = -0.8306, 95\% CI = [0.4047, 0.5175], p-value = 0.0056\\
RH Cont: r = -0.9353, 95\% CI = [0.5334, 0.5971], p-value = 0.0002\\
RH Default: r = -0.9605, 95\% CI = [0.5522, 0.6330], p-value = 0.0000\\
\section*{SI.7: Regression of DVARS for different parcellation scales (ROI = 100, 200, 300, and 400, PA)}
\captionsetup[figure]{labelfont={bf},name={Fig },labelsep=period}
\renewcommand{\thefigure}{SI.7}
\begin{figure*}[!ht]
\centering
\includegraphics[width=0.9\linewidth]{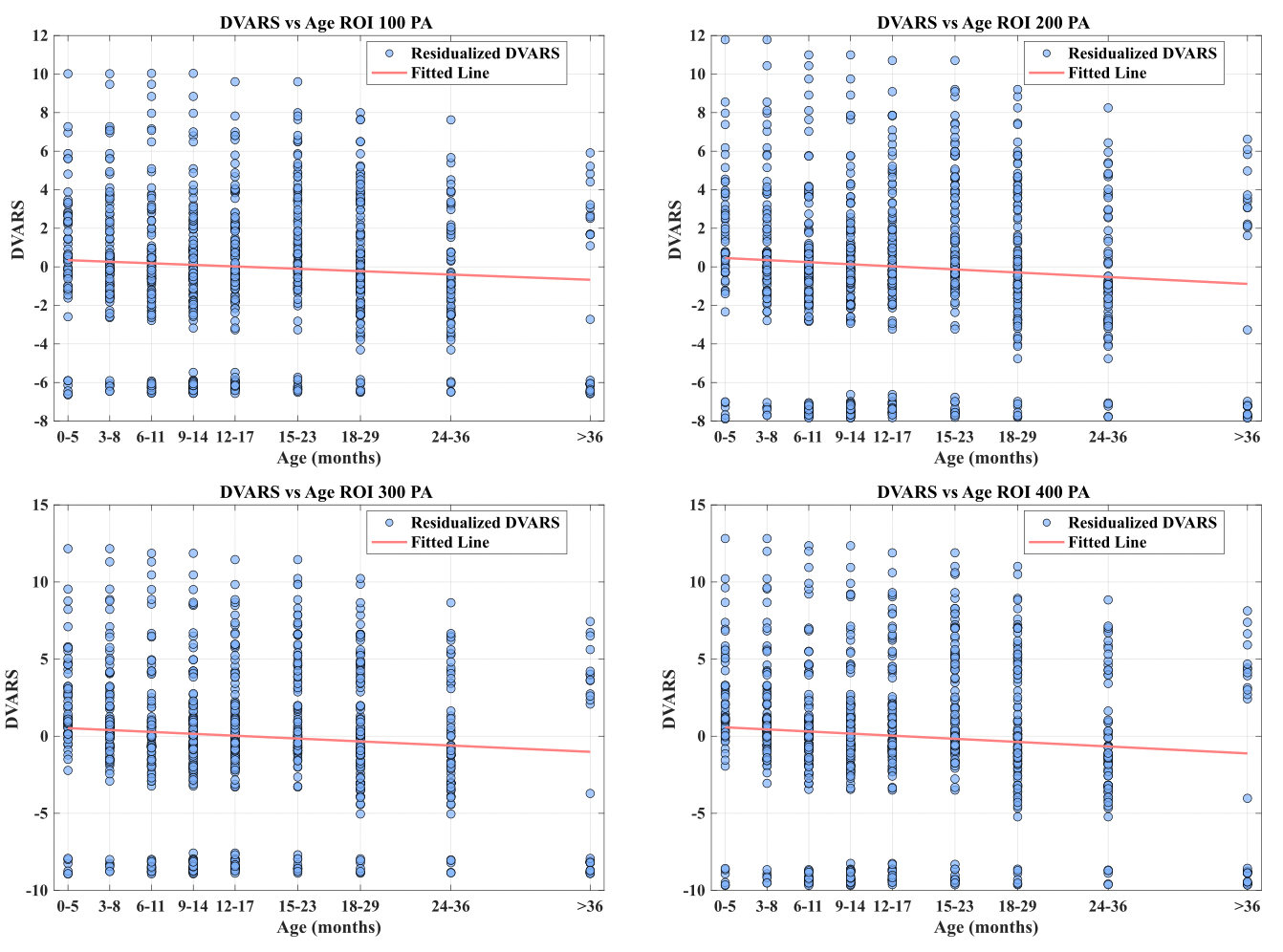}
\caption{\footnotesize \textbf{}}
\label{SI_7}
\end{figure*}
\textbf{ }\\

\section*{SI.8: SFMC with alternative age grouping 0-4, 3-7 etc}
\captionsetup[table]{labelfont={bf},name={Table },labelsep=period}
\renewcommand{\thetable}{SI.8}
\begin{table}[!ht]
\caption{Age ranges with different numbers of scans for fMRI and dMRI.}
\centering
\begin{tabular}{|c|c|c|c|c|c|c|c|c|c|c|r|l|} \hline
 Age ranges (month) & 0-4 & 3-7 & 5-9 & 8-12 & 10-14 & 13-17 & 15-19 & 17-26 & 20-36 & $>$36  \\ \hline
fMRI Scans (AP) & 41 & 73 & 90 & 108 & 99 & 88 & 80 & 112 & 94 & 29\\ \hline
fMRI Scans (PA) & 41 & 74 & 92 & 113 & 103 & 87 & 76 & 115 & 97 & 29 \\ \hline
dMRI Scans (AP) & 37 & 80 & 99 & 99 & 84 & 77 & 68 & 107 & 87 & 16 \\ \hline
dMRI Scans (PA) & 37 & 80 & 99 & 99 & 84 & 77 & 68 & 107 & 87 & 16\\ \hline
\end{tabular}
\label{agewindows}
\end{table}

\captionsetup[figure]{labelfont={bf},name={Fig },labelsep=period}
\renewcommand{\thefigure}{SI.8-1}
\begin{figure*}[!ht]
\centering
\includegraphics[width=0.9\linewidth]{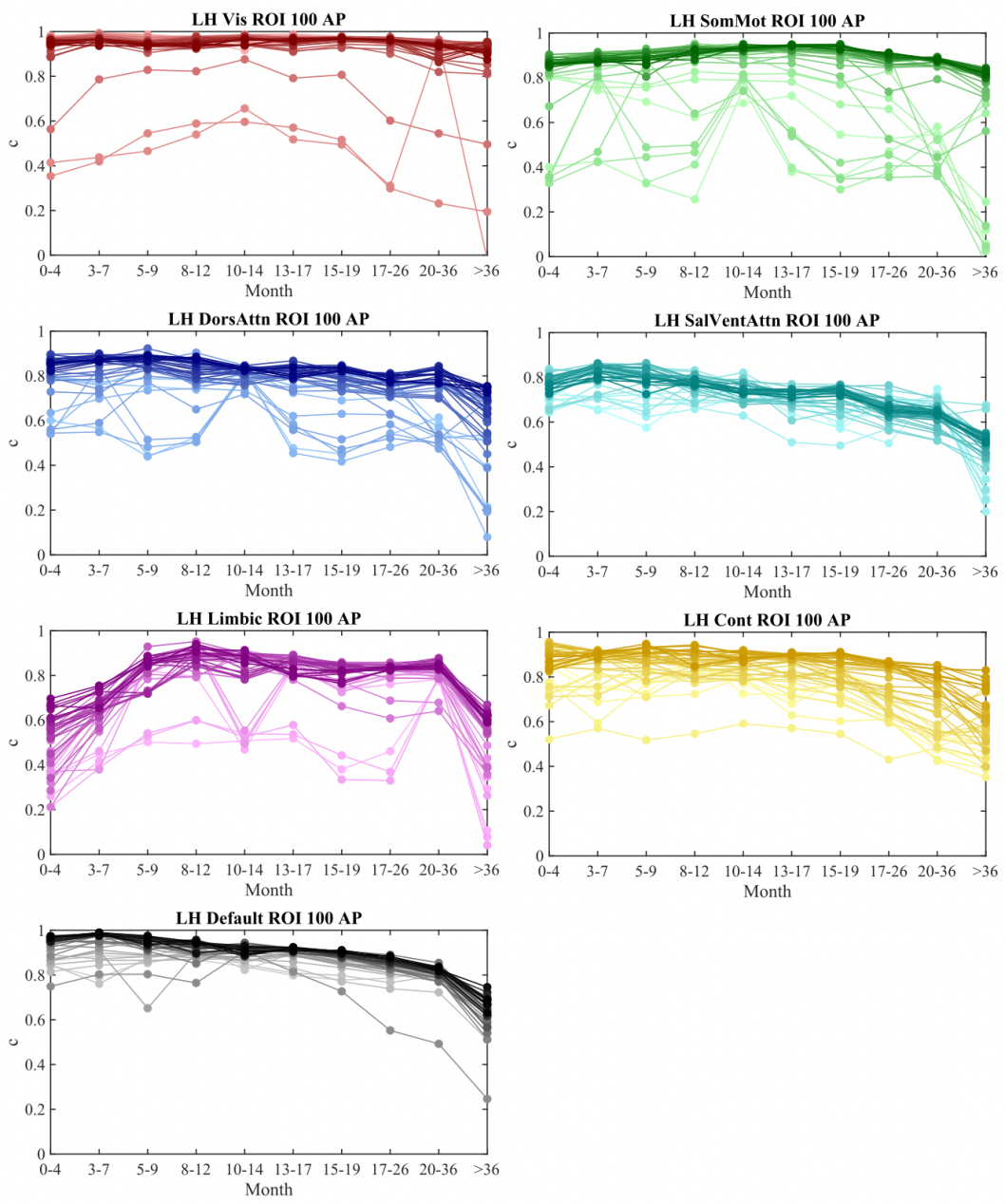}
\caption{\footnotesize \textbf{}}
\label{SI_8_1}
\end{figure*}
\textbf{ }\\

\captionsetup[figure]{labelfont={bf},name={Fig },labelsep=period}
\renewcommand{\thefigure}{SI.8-2}
\begin{figure*}[!ht]
\centering
\includegraphics[width=0.9\linewidth]{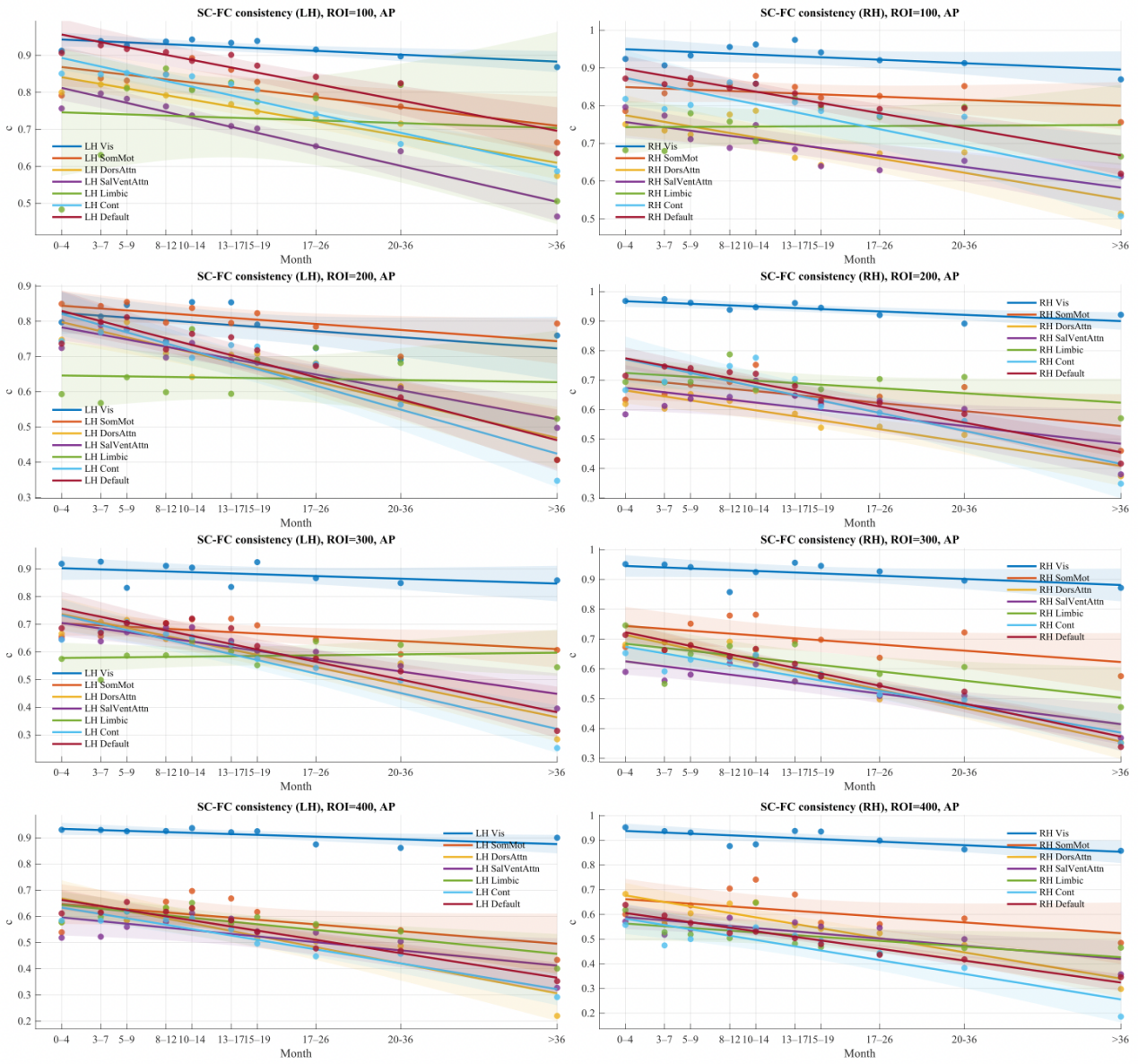}
\caption{\footnotesize \textbf{}}
\label{SI_8_2}
\end{figure*}
\textbf{ }\\
\\
\textbf{ROI=100, AP}\\
LH Vis: r = -0.7680, 95\% CI = [0.9240, 0.9616], p-value = 0.0095\\
LH SomMot: r = -0.7549, 95\% CI = [0.8164, 0.9203], p-value = 0.0116\\
LH DorsAttn: r = -0.9482, 95\% CI = [0.8117, 0.8702], p-value = 0.0000\\
LH SalVentAttn: r = -0.9594, 95\% CI = [0.7779, 0.8462], p-value = 0.0000\\
LH Limbic: r = -0.0929, 95\% CI = [0.5744, 0.9177], p-value = 0.7985\\
LH Cont: r = -0.9597, 95\% CI = [0.8604, 0.9256], p-value = 0.0000\\
LH Default: r = -0.9191, 95\% CI = [0.9141, 0.9983], p-value = 0.0002\\
RH Vis: r = -0.5370, 95\% CI = [0.9176, 0.9815], p-value = 0.1094\\
RH SomMot: r = -0.4099, 95\% CI = [0.8078, 0.8909], p-value = 0.2394\\
RH DorsAttn: r = -0.8433, 95\% CI = [0.7210, 0.8280], p-value = 0.0022\\
RH SalVentAttn: r = -0.8385, 95\% CI = [0.7139, 0.7990], p-value = 0.0024\\
RH Limbic: r = 0.0313, 95\% CI = [0.6729, 0.8132], p-value = 0.9316\\
RH Cont: r = -0.8045, 95\% CI = [0.7997, 0.9473], p-value = 0.0050\\
RH Default: r = -0.9275, 95\% CI = [0.8627, 0.9325], p-value = 0.0001\\
\\
\textbf{ROI=200, AP}\\
LH Vis: r = -0.5393, 95\% CI = [0.7644, 0.8833], p-value = 0.1077\\
LH SomMot: r = -0.6658, 95\% CI = [0.8016, 0.8867], p-value = 0.0356\\
LH DorsAttn: r = -0.9048, 95\% CI = [0.7393, 0.8561], p-value = 0.0003\\
LH SalVentAttn: r = -0.9359, 95\% CI = [0.7457, 0.8198], p-value = 0.0001\\
LH Limbic: r = -0.0740, 95\% CI = [0.5492, 0.7431], p-value = 0.8390\\
LH Cont: r = -0.9217, 95\% CI = [0.7579, 0.8839], p-value = 0.0001\\
LH Default: r = -0.9231, 95\% CI = [0.7717, 0.8871], p-value = 0.0001\\
RH Vis: r = -0.7953, 95\% CI = [0.9484, 0.9873], p-value = 0.0060\\
RH SomMot: r = -0.6472, 95\% CI = [0.6336, 0.7762], p-value = 0.0431\\
RH DorsAttn: r = -0.9081, 95\% CI = [0.6205, 0.7096], p-value = 0.0003\\
RH SalVentAttn: r = -0.6796, 95\% CI = [0.5967, 0.7510], p-value = 0.0306\\
RH Limbic: r = -0.5775, 95\% CI = [0.6707, 0.7782], p-value = 0.0804\\
RH Cont: r = -0.8670, 95\% CI = [0.6941, 0.8483], p-value = 0.0012\\
RH Default: r = -0.9560, 95\% CI = [0.7372, 0.8111], p-value = 0.0000\\
\\
\textbf{ROI=300, AP}\\
LH Vis: r = -0.4436, 95\% CI = [0.8606, 0.9455], p-value = 0.1991\\
LH SomMot: r = -0.6198, 95\% CI = [0.6591, 0.7482], p-value = 0.0560\\
LH DorsAttn: r = -0.9247, 95\% CI = [0.6786, 0.7945], p-value = 0.0001\\
LH SalVentAttn: r = -0.8916, 95\% CI = [0.6560, 0.7543], p-value = 0.0005\\
LH Limbic: r = 0.1284, 95\% CI = [0.5230, 0.6334], p-value = 0.7237\\
LH Cont: r = -0.9387, 95\% CI = [0.6759, 0.7899], p-value = 0.0001\\
LH Default: r = -0.9178, 95\% CI = [0.6956, 0.8179], p-value = 0.0002\\
RH Vis: r = -0.5518, 95\% CI = [0.9089, 0.9816], p-value = 0.0982\\
RH SomMot: r = -0.5787, 95\% CI = [0.6796, 0.8082], p-value = 0.0796\\
RH DorsAttn: r = -0.9615, 95\% CI = [0.6751, 0.7521], p-value = 0.0000\\
RH SalVentAttn: r = -0.8716, 95\% CI = [0.5806, 0.6701], p-value = 0.0010\\
RH Limbic: r = -0.7086, 95\% CI = [0.6166, 0.7527], p-value = 0.0218\\
RH Cont: r = -0.9358, 95\% CI = [0.6335, 0.7154], p-value = 0.0001\\
RH Default: r = -0.9733, 95\% CI = [0.6918, 0.7540], p-value = 0.0000\\
\\
\textbf{ROI=400, AP}\\
LH Vis: r = -0.6873, 95\% CI = [0.9110, 0.9575], p-value = 0.0281\\
LH SomMot: r = -0.6010, 95\% CI = [0.5711, 0.7228], p-value = 0.0661\\
LH DorsAttn: r = -0.8913, 95\% CI = [0.5988, 0.7379], p-value = 0.0005\\
LH SalVentAttn: r = -0.7096, 95\% CI = [0.5272, 0.6652], p-value = 0.0215\\
LH Limbic: r = -0.8143, 95\% CI = [0.5936, 0.6939], p-value = 0.0041\\
LH Cont: r = -0.9457, 95\% CI = [0.5934, 0.6742], p-value = 0.0000\\
LH Default: r = -0.9458, 95\% CI = [0.6244, 0.7013], p-value = 0.0000\\
RH Vis: r = -0.7272, 95\% CI = [0.9080, 0.9682], p-value = 0.0172\\
RH SomMot: r = -0.5350, 95\% CI = [0.5797, 0.7438], p-value = 0.1110\\
RH DorsAttn: r = -0.9288, 95\% CI = [0.6261, 0.7275], p-value = 0.0001\\
RH SalVentAttn: r = -0.7941, 95\% CI = [0.5403, 0.6384], p-value = 0.0061\\
RH Limbic: r = -0.6056, 95\% CI = [0.4954, 0.6305], p-value = 0.0635\\
RH Cont: r = -0.8947, 95\% CI = [0.5212, 0.6444], p-value = 0.0005\\
RH Default: r = -0.9773, 95\% CI = [0.5826, 0.6287], p-value = 0.0000\\

\section*{SI.9: SFMC with alternative age grouping 0-6, 3-9 etc}
\renewcommand{\thetable}{SI.9}
\begin{table}[!ht]
\caption{Age ranges with different numbers of scans for fMRI and dMRI.}
\centering
\begin{tabular}{|c|c|c|c|c|c|c|c|c|c|r|l|} \hline
 Age ranges (month) & 0-6 & 3-9 & 7-13 & 10-16 & 14-20 & 17-26 & 21-30 & 27-36 & $>$36  \\ \hline
fMRI Scans (AP)     & 74 & 117 & 138 & 141 & 104 & 112 & 75 & 20 & 29  \\ \hline
fMRI Scans (PA)     & 76 & 118 & 141 & 146 & 103 & 115 & 78 & 15 & 29   \\ \hline
dMRI Scans (AP)     & 77 & 126 & 126 & 120 & 92 & 107 & 73 & 12 & 16   \\ \hline
dMRI Scans (PA)     & 77 & 126 & 126 & 120 & 92 & 107 & 73 & 12 & 16  \\ \hline
\end{tabular}
\label{agewindows}
\end{table}

\captionsetup[figure]{labelfont={bf},name={Fig },labelsep=period}
\renewcommand{\thefigure}{SI.9-1}
\begin{figure*}[!ht]
\centering
\includegraphics[width=0.9\linewidth]{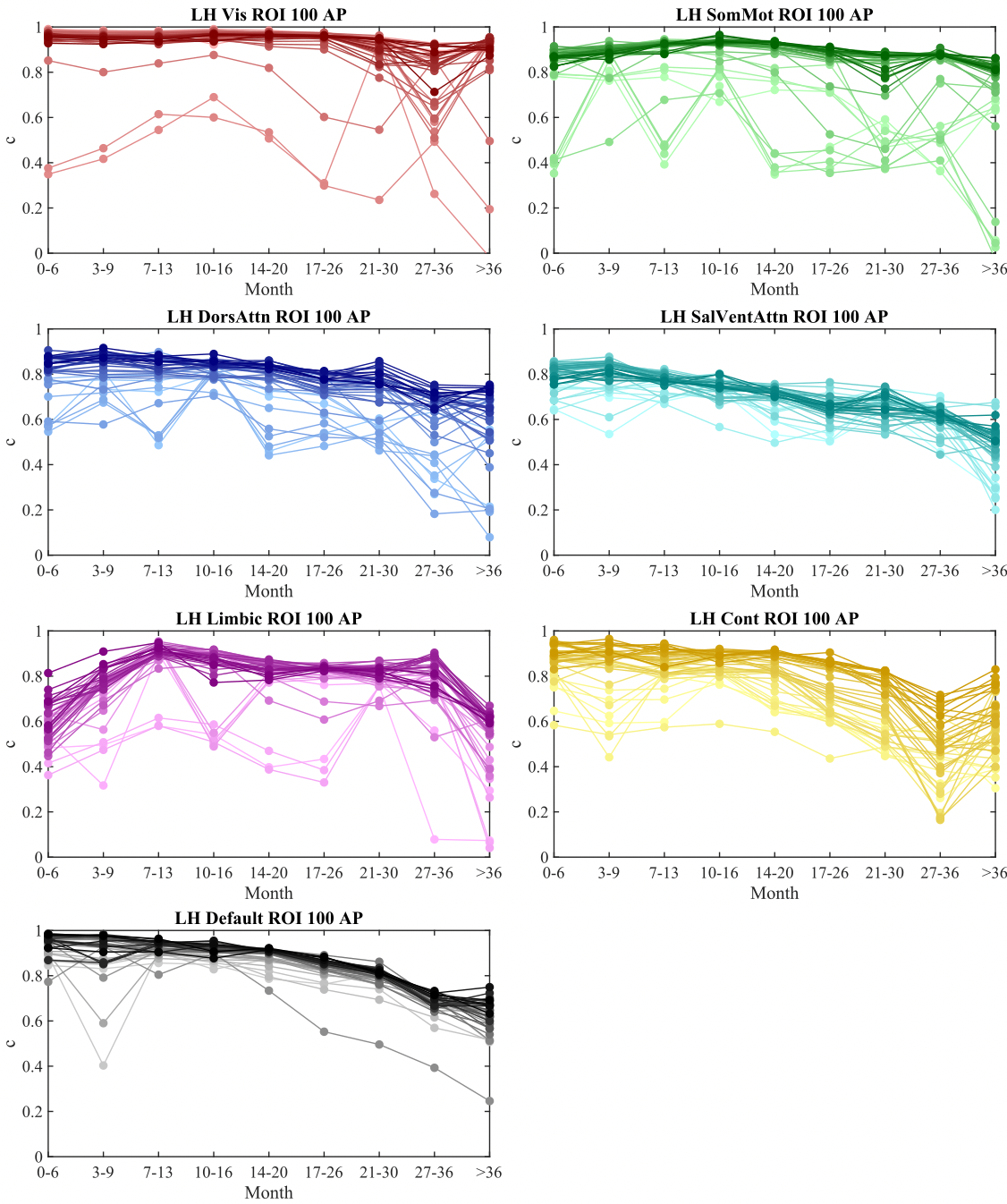}
\caption{\footnotesize \textbf{}}
\label{SI_9_1}
\end{figure*}
\textbf{ }\\

\captionsetup[figure]{labelfont={bf},name={Fig },labelsep=period}
\renewcommand{\thefigure}{SI.9-2}
\begin{figure*}[!ht]
\centering
\includegraphics[width=0.9\linewidth]{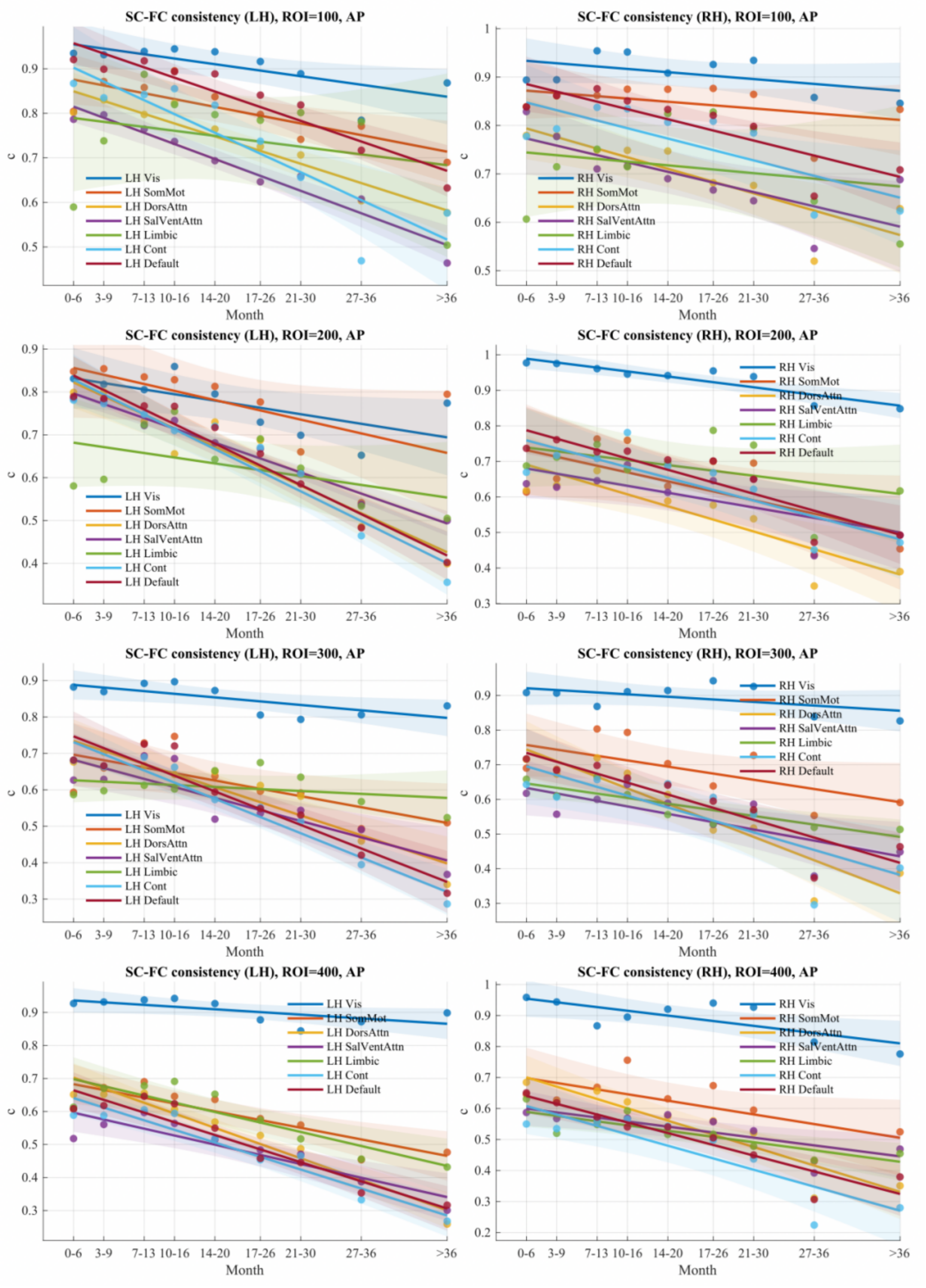}
\caption{\footnotesize \textbf{}}
\label{SI_9_1}
\end{figure*}
\textbf{ }\\
\\
\textbf{ROI=100, AP}\\
LH Vis: r = -0.7453, 95\% CI = [0.9054, 1.0039], p-value = 0.0212\\
LH SomMot: r = -0.8074, 95\% CI = [0.8202, 0.9309], p-value = 0.0085\\
LH DorsAttn: r = -0.9498, 95\% CI = [0.8082, 0.8907], p-value = 0.0001\\
LH SalVentAttn: r = -0.9721, 95\% CI = [0.7799, 0.8501], p-value = 0.0000\\
LH Limbic: r = -0.2893, 95\% CI = [0.6251, 0.9538], p-value = 0.4503\\
LH Cont: r = -0.8932, 95\% CI = [0.8113, 0.9934], p-value = 0.0012\\
LH Default: r = -0.9459, 95\% CI = [0.9113, 1.0033], p-value = 0.0001\\
RH Vis: r = -0.5312, 95\% CI = [0.8873, 0.9799], p-value = 0.1411\\
RH SomMot: r = -0.4392, 95\% CI = [0.8140, 0.9301], p-value = 0.2369\\
RH DorsAttn: r = -0.8521, 95\% CI = [0.7306, 0.8573], p-value = 0.0035\\
RH SalVentAttn: r = -0.7529, 95\% CI = [0.6987, 0.8479], p-value = 0.0192\\
RH Limbic: r = -0.2402, 95\% CI = [0.6112, 0.8774], p-value = 0.5335\\
RH Cont: r = -0.7644, 95\% CI = [0.7698, 0.9252], p-value = 0.0165\\
RH Default: r = -0.8454, 95\% CI = [0.8288, 0.9423], p-value = 0.0041\\
\\
\textbf{ROI=200, AP}\\
LH Vis: r = -0.6752, 95\% CI = [0.7614, 0.9025], p-value = 0.0460\\
LH SomMot: r = -0.6258, 95\% CI = [0.7403, 0.9719], p-value = 0.0714\\
LH DorsAttn: r = -0.9521, 95\% CI = [0.7626, 0.8818], p-value = 0.0001\\
LH SalVentAttn: r = -0.9873, 95\% CI = [0.7736, 0.8193], p-value = 0.0000\\
LH Limbic: r = -0.5021, 95\% CI = [0.5784, 0.7851], p-value = 0.1684\\
LH Cont: r = -0.9600, 95\% CI = [0.7702, 0.8872], p-value = 0.0000\\
LH Default: r = -0.9747, 95\% CI = [0.7933, 0.8836], p-value = 0.0000\\
RH Vis: r = -0.9099, 95\% CI = [0.9608, 1.0173], p-value = 0.0007\\
RH SomMot: r = -0.6467, 95\% CI = [0.6037, 0.8602], p-value = 0.0598\\
RH DorsAttn: r = -0.8703, 95\% CI = [0.6090, 0.7731], p-value = 0.0023\\
RH SalVentAttn: r = -0.7110, 95\% CI = [0.5966, 0.7638], p-value = 0.0317\\
RH Limbic: r = -0.4837, 95\% CI = [0.6285, 0.8499], p-value = 0.1871\\
RH Cont: r = -0.8273, 95\% CI = [0.6707, 0.8479], p-value = 0.0059\\
RH Default: r = -0.9026, 95\% CI = [0.7221, 0.8534], p-value = 0.0009\\
\\
\textbf{ROI=300, AP}\\
LH Vis: r = -0.7294, 95\% CI = [0.8482, 0.9278], p-value = 0.0257\\
LH SomMot: r = -0.7129, 95\% CI = [0.6102, 0.7828], p-value = 0.0311\\
LH DorsAttn: r = -0.9372, 95\% CI = [0.6757, 0.7930], p-value = 0.0002\\
LH SalVentAttn: r = -0.8743, 95\% CI = [0.6104, 0.7540], p-value = 0.0020\\
LH Limbic: r = -0.3525, 95\% CI = [0.5659, 0.6865], p-value = 0.3522\\
LH Cont: r = -0.9675, 95\% CI = [0.6806, 0.7812], p-value = 0.0000\\
LH Default: r = -0.9401, 95\% CI = [0.6789, 0.8146], p-value = 0.0002\\
RH Vis: r = -0.5366, 95\% CI = [0.8731, 0.9687], p-value = 0.1364\\
RH SomMot: r = -0.6499, 95\% CI = [0.6668, 0.8476], p-value = 0.0581\\
RH DorsAttn: r = -0.9225, 95\% CI = [0.6628, 0.8254], p-value = 0.0004\\
RH SalVentAttn: r = -0.7598, 95\% CI = [0.5541, 0.7124], p-value = 0.0175\\
RH Limbic: r = -0.8751, 95\% CI = [0.6061, 0.6859], p-value = 0.0020\\
RH Cont: r = -0.8044, 95\% CI = [0.5871, 0.8039], p-value = 0.0090\\
RH Default: r = -0.9053, 95\% CI = [0.6646, 0.8041], p-value = 0.0008\\
\\
\textbf{ROI=400, AP}\\
LH Vis: r = -0.6681, 95\% CI = [0.8995, 0.9733], p-value = 0.0492\\
LH SomMot: r = -0.8640, 95\% CI = [0.6235, 0.7421], p-value = 0.0027\\
LH DorsAttn: r = -0.9742, 95\% CI = [0.6589, 0.7460], p-value = 0.0000\\
LH SalVentAttn: r = -0.9012, 95\% CI = [0.5388, 0.6540], p-value = 0.0009\\
LH Limbic: r = -0.8815, 95\% CI = [0.6320, 0.7639], p-value = 0.0017\\
LH Cont: r = -0.9569, 95\% CI = [0.5900, 0.6912], p-value = 0.0001\\
LH Default: r = -0.9520, 95\% CI = [0.6104, 0.7183], p-value = 0.0001\\
RH Vis: r = -0.7573, 95\% CI = [0.8962, 1.0127], p-value = 0.0181\\
RH SomMot: r = -0.6775, 95\% CI = [0.6003, 0.7962], p-value = 0.0449\\
RH DorsAttn: r = -0.9275, 95\% CI = [0.6310, 0.7704], p-value = 0.0003\\
RH SalVentAttn: r = -0.7915, 95\% CI = [0.5443, 0.6560], p-value = 0.0110\\
RH Limbic: r = -0.8419, 95\% CI = [0.5410, 0.6378], p-value = 0.0044\\
RH Cont: r = -0.8694, 95\% CI = [0.5182, 0.6979], p-value = 0.0023\\
RH Default: r = -0.9349, 95\% CI = [0.5845, 0.6969], p-value = 0.0002\\

\end{document}